# Technical debt and agile software development practices and processes: An industry practitioner survey

Johannes Holvitie[✉,1,2], Sherlock A. Licorish[3], Rodrigo O. Spínola[4,5], Sami Hyrynsalmi[6], Stephen G. MacDonell[3,8], Thiago S. Mendes[7], Jim Buchan[8], Ville Leppänen[1,2]

[1]*Turku Centre for Computer Science, Software Development Laboratory, Turku, Finland*
[2]*University of Turku, Department of Information Technology, Turku, Finland*
[3]*University of Otago, Department of Information Science, Dunedin, Otago, New Zealand*
[4]*Salvador University, Graduate Program in Systems and Computer, Salvador, Bahia, Brazil*
[5]*Federal University of Bahia, Fraunhofer Project Center for Software and System Engineering, Salvador, Brazil*
[6]*Tampere University of Technology, Pori Campus, Pori, Finland*
[7]*Federal Institute of Bahia, Information Technology Department, Santo Amaro, Bahia, Brazil*
[8]*Auckland University of Technology, School of Engineering, Computer and Mathematical Sciences, Auckland, New Zealand*

**Abstract**

***Context:*** *Contemporary software development is typically conducted in dynamic, resource-scarce environments that are prone to the accumulation of technical debt. While this general phenomenon is acknowledged, what remains unknown is how technical debt specifically manifests in and affects software processes, and how the software development techniques employed accommodate or mitigate the presence of this debt.* ***Objectives:*** *We sought to draw on practitioner insights and experiences in order to classify the effects of agile method use on technical debt management, given the popularity and perceived success of agile methods. We explore the breadth of practitioners' knowledge about technical debt; how technical debt is manifested across the software process; and the perceived effects of common agile software development practices and processes on technical debt. In doing so, we address a research gap in technical debt knowledge and provide novel and actionable managerial recommendations.* ***Method:*** *We designed, tested and executed a multi-national survey questionnaire to address our objectives, receiving 184 responses from practitioners in Brazil, Finland, and New Zealand.* ***Results:*** *Our findings indicate that: 1) Practitioners are aware of technical debt, although, there was under utilization of the concept, 2) Technical debt commonly resides in legacy systems, however, concrete instances of technical debt are hard to conceptualize which makes it problematic to manage, 3) Queried agile practices and processes help to reduce technical debt; in particular, techniques that verify and maintain the structure and clarity of implemented artifacts (e.g., Coding standards and Refactoring) positively affect technical debt management.* ***Conclusions:*** *The fact that technical debt instances tend to have characteristics in common means that a systematic approach to its management is feasible. However, notwithstanding the positive effects of some agile practices on technical debt management, competing stakeholders' interests remain a concern.*

## 1. INTRODUCTION

A software developer can follow several alternative paths to produce "working software" [1]. Arguably, these paths influence, and are influenced by, stakeholders' definitions of done [2]. All of the alternatives may produce working software for a client, but in taking some paths the resulting implementation may be more complex than those arising from other approaches. If there are to be future iterations in the development of a software product or service, and these iterations must build on a more complex implementation, it is evident that the chosen path affects and will continue to affect development: it is easier to progress software development on top of a less complex implementation.

The potential misalignment of client and developer perspectives regarding the criteria for judging whether or not a feature is done can lead to technical debt. The client reviews the functional completeness of features to inform decisions regarding later iterations. The developer, however, observes both functional and implementation completeness for the features. As implementation-related matters are largely invisible to the client the developer is primarily responsible for managing completeness in this dimension. Notably, the developer must understand how that completeness affects delivery to the client and, based on this understanding, exercise appropriate management. For example, say that in an upcoming iteration a component for which an ad hoc interface was designed will be used. The developer is the only party capable of knowing this and so must decide whether spending part of the iteration on an interface redesign is a preferred strategy over using the existing, incoherent interface. Thus, the dual perspectives on the definition of done, along with perceptions of the gap from what can be observed by all stakeholders and software feature completeness, and the foreseeable actions or paths that could be taken regarding that gap, provide the reasoning for technical debt's existence, describe its manifestation in practice, and lead to its management procedures, respectively.

While as a phenomenon technical debt is not new, its conceptualization is quite recent [3]. Technical debt describes the consequences of software development actions that intentionally or unintentionally prioritize client value and/or project constraints such as delivery deadlines, over more technical implementation and design considerations. These include matters such as achieving and sustaining test coverage or code extensibility. Conceptually, technical debt is an analog of financial debt, with associated concepts such as levels of debt, debt accrual over time and its likely consequences, and the pressure to pay back the debt at some point in time.

Technical debt should not be equated with sub-optimal software and the negative effects arising from such developments, however. There are circumstances when the decision to accrue technical debt (i.e., not pay it back) has positive cost-benefit to a team. The use of technical debt ideas to compartmentalize and characterize the deviation between current and optimal software states can provide a mechanism for asset-management-like governance of the debt [4]. For example, a decision to not spend resources on improving a working software structure that delivers the desired functionality is reasonable if available information indicates that there is likely to be no (timely) advantage or added benefit in return for this effort.

While the consideration and application of the technical debt concept have increased exponentially in the academic context [5], to the best of the authors' knowledge several aspects of the concept's use in the software industry remain unstudied, including the contexts within which the effects of technical debt are likely to have the greatest relevance and impact. In particular, few prior studies have attempted to capture the effects of common software development practices and processes - the paths taken - on technical debt. In the same vein, as a community we are unsure about the breadth of practitioners' knowledge about technical debt and how technical debt is manifested as they work. The word 'effect' is used here to capture if a particular practice or process is perceived to increase or decrease the size of technical debt or the positive or negative outcomes that emerge from this debt.

Knowledge pertaining to the effects of software development practices or processes on technical debt is potentially important to practitioners since they must make decisions about the development methods they will use. In the absence of sufficient knowledge about the effects of a method's practices and processes, these decisions may have unintended consequences on technical debt, and ultimately, teams' performance. Based on such observations, and in keeping with our desire to gain a broad range of input to underpin our understanding, we were prompted to conduct an multi-national survey to investigate technical debt in practice. This exploratory study sought to extend our knowledge of the depth and breadth of practitioners' knowledge about technical debt; how technical debt is manifested; and the perceived effects of common agile software development practices and processes on technical debt.

This work significantly extends the authors' prior preliminary contributions. A previous publication involving some of the present authors [6] reports the design, construction, and testing of the survey instrument, in addition to its execution in Finland. That previous publication describes scenarios where technical debt is used, the media in which respondents use the concept, and their existing knowledge levels. Further, the perceived effect of common agile practices and processes on technical debt is queried. The effect of continued development, the causes, and the origins are also captured for technical debt.

The current work extends this initial study to a multi-national one that involves participants in Brazil and New Zealand. This enabled the delivery of several novel results. Most notably, analysis of previous knowledge, conformity to given definitions and agile technique effects now consider the respondents' development roles. We also report on technical debt's effects on several software development characteristics (e.g., its perceived effect on agility). Further, due to the increased dataset size the analysis confidence is increased, and so where applicable, statistical analysis accompanies the presentation of results.

The remaining sections of this article are structured as follows. Section 2 describes the background of the study, focusing on related work on technical debt and agile software development. The research approach employed is described in Section 3 with the establishment of the research questions, followed by explanations of the design and implementation of the survey study that answers them. Section 4 presents the results of the survey and the subsequent analysis of these results. Key findings, implications and future work, as well as study limitations, are discussed in Section 5. Finally, concluding remarks are provided in Section 6.

## 2. BACKGROUND

In this section we provide the study background. We first examine the concept of technical debt, including its origins and evolution. We then provide a review and evaluation of existing relevant cross-sectional studies, noting the significant characteristics of technical debt. This review and evaluation provide the working definition of technical debt used in our study. Finally, agile software development and its methods are briefly examined so as to provide context and justification for our consideration of particular agile practices and processes in the survey instrument.

**2.1. Defining technical debt**
The term "technical debt" was coined by Ward Cunningham [3], when he described the phenomenon of meeting a release deadline by making adaptations and concessions in a product. He also outlined how the effects felt afterwards were analogous to those associated with the incurring of financial debt. Cunningham [3] acknowledged that, most often, technical debt required payback, while the inability to manage assets could lead to a complete stand-still as the interest and effects of the adaptations (or lack thereof) become unbearable.

The definition of technical debt was later revisited on a number of occasions, usually to generalize what Cunningham had previously described for all applicable situations whilst categorizing its characteristics. Steve McConnell's definition, which separates out intentional and unintentional accumulation of technical debt [7], has been widely adopted by academia (e.g., in [8,9] and recognized in [5]):

> The first kind of technical debt is the kind that is incurred unintentionally. For example, a design approach just turns out to be error-prone or a junior programmer just writes bad code. This technical debt is the non-strategic result of doing a poor job. In some cases, this kind of debt can be incurred unknowingly...

> The second kind of technical debt is the kind that is incurred intentionally. This commonly occurs when an organization makes a conscious decision to optimize for the present rather than for the future. "If we don't get this release done on time, there won't be a next release"...

Building on McConnell's assessment, Brown et al. [10] provide a description of the effects of technical debt during software development by relating the concept to its financial counterpart:

> The metaphor highlights that, like financial debt, technical debt incurs interest payments in the form of increased future costs owing to earlier quick and dirty design and implementation choices. Like financial debt, sometimes technical debt can be necessary. One can continue paying interest, or pay down the principal by rearchitecting and refactoring to reduce future interest payments.

As this definition is the sole definition—to the best of the authors' knowledge—with an explicit monetary reference, it is of interest from a definition perspective when engaging practitioners (see the Questionnaire for operationalization of both definitions http://soft.utu.fi/tds16/questionnaire.pdf , Question 22).

Given that the previously synthesized definitions are both abstract and generic, other researchers have sought to contextualize technical debt. For instance, Alves et al. [9] provided an ontology for technical debt, explaining that 13 different types of technical debt can be usefully distinguished (e.g., design, architecture, and testing debt [11,12]). The contextualization of technical debt in their model is apparent through the type name which indicates the software context from which the type's instances originate. Kruchten et al. have provided a Technical Debt Landscape [8] (further discussed by Izurieta et al. [13]) which captures a set of causes that are said to lead to the emergence of technical debt.

In the Technical Debt Landscape, the causes are introduced by placing them relative to two axes. The first axis characterizes their visibility, from visible (i.e., new features, additional functionality, defects, and low external quality) to mostly invisible (i.e., architectural debt structural debt, test debt, documentation debt, code complexity, low internal quality, coding style violations, and code smells) while the second axis characterizes the issue type, from evolution to quality (where of the previously listed issues, new features, additional functionality, architectural debt, structural debt, and test debt are evolution issues, coding style violations, code smells, defects, and low external quality are quality issues, and documentation debt, code complexity, and low internal quality fall in between).

Authors of the landscape argue that causes that are clearly visible in fact do not count as technical debt while the mostly invisible issues do count. The reasoning given for this division is that the visible issues have explicit and self-evident representations. The development organization hence understands the need, or at least the opportunity, to manage them. The mostly invisible issues accumulate technical debt because their emergence, presence, and negative effects remain largely unknown. Only software developers who have gained additional knowledge through development understand these issues.

In our survey instrument we have sought to consider both these visible and invisible causes as we want to discover if respondents similarly perceive this division (see http://soft.utu.fi/tds16/ questionnaire.pdf , Question 31 for the authors' adaptation of this).

In going one step further, a number of studies have endeavoured to validate various contextualizations and clarifications of technical debt, often through the use of interview or survey instruments. We review these next to situate our study in the current state of knowledge.

**2.2. Cross-sectional studies of technical debt**

A wide range of cross-sectional studies of technical debt have been reported in the literature, some of which sought to characterize the phenomenon while others addressed how technical debt is incurred. We
review a non-exhaustive selection of those studies here in order to provide the context for our own survey-based work.

Klinger et al. [14] interviewed four software architects concerning technical debt accumulation. They noted that often the decision to incur debt was derived from the motivations of non-technical stakeholders, and hence, could be affected by competing stakeholder interests (e.g., a technically sub-optimal solution was chosen due to pressing business concerns). It was nevertheless concluded that quantification of technical debt should make relevant information more available (so as to be visible to the project team), and thus, should enable more effective project management.

Snipes et al. [15] surveyed two change control boards to understand the decision factors surrounding defect debt governance. They identified a set of factors that affected accumulation and reduction of technical debt by relating these to the success of chosen management strategies. Snipes et al. [15] further noted that most of the software defect management resources were spent on identification and characterization, rather than the actual removal of defects (and ultimately, the reduction of technical debt).

Martini and Bosch [16] studied the information needs of agile software architects and product owners with regard to architectural technical debt (ATD). They analyzed quantitative and qualitative data collected from four large software development companies. Their findings indicated that, for architect and product owner roles, the information needs were different. Notably, product owners valued market attractiveness and specific customer value more highly than the software architects.

Martini et al. [17] also reported on a study to establish causes for the accumulation of ATD, as well as its refactoring. The study was executed as a multiple-case study in five large software development companies. Based on the executed interviews a series of causes for ATD was identified and validated (e.g., uncertainty of use cases in early stages). Martini et al. presented two models for ATD accumulation and evolution: crisis and phases. The Crisis Model captured a contemporary phenomenon where ATD was allowed to accumulate up until the point when adding new business value was too cumbersome, and a large refactoring had to be executed. The Phases Model captured the different time periods in relation to feature research, design, and implementation, that could be identified to have differing ATD accumulation properties. Hence this model could be used to acknowledge differing ATD accumulation, avoidance and refactoring opportunities.

Spinola et al. [18] compiled a list of technical debt folklore and surveyed practitioners to ascertain the extent of their agreement with those beliefs. Consensus was indicated for: *"technical debt often accumulates via short-term optimizations, reduction of technical debt is good for morale, non-management of technical debt results in unsustainability"*, and *"not all technical debt is negative - which is why it should not be avoided, but rather, managed"*. The study's authors noted, however, that the low number of responses ($N = 37$) limited the generalizability of their results.

Lim et al. [19] interviewed 35 practitioners regarding how technical debt manifested and how it was generally managed. In their study, Lim et al. observed that 75% of their respondents initially indicated that they were not familiar with the term "technical debt". After describing the concept, those respondents who were familiar with the concept indicated that informed decisions sometimes resulted in the team incurring technical debt, and that the effects of this phenomenon were long-term. Further, management of technical debt was seen to be generally difficult, as tracking non-uniform technical debt instances was a challenging exercise.

Codabux et al. [20] studied, within a large software organization, how technical debt could be characterized, the debt's effects on software development, and its management procedures. They concluded that an in-house taxonomy was perceived as useful for technical debt characterization, while explicit measures were encouraged for management (e.g., dedicated teams and task descriptions). Codabux et al. [20] found that prioritizing the management of technical debt, particularly based on stakeholders' perceptions of debt severity, was ranked high among the measures for countering technical debt. These results were gathered through interviews with 28 project managers.

Ernst et al. [21] surveyed 1831 respondents from three large organizations and received 536 fully completed questionnaires. They explored whether practitioners shared a common technical debt definition, if issues with architectural components were among the main contributors of technical debt, and if there were practices and tools readily available for technical debt management and tracking. These authors established that software practitioners and managers had a uniform understanding of technical debt. In addition, the study's authors found that architectural choices, especially in early stages of the software development life-cycle (as noted also by Martini et al. [17]), were a major contributor of technical debt. Further, it was revealed that there is a lack of tool support for managing architectural technical debt. However, there are limits on the degree to which the results of Ernst et al. could be considered as general or typical, given they captured perceptions from employees at three companies.

Finally, a very recent study by Behutiye et al. [22] reports the results of a systematic literature review that addressed the concept of technical debt in the agile software development context. The analysis of 38 relevant articles led the authors to pinpoint research areas of interest, technical debt cause and effect classifications, as well as management strategies from existing research. The outcomes of this work add to the body of knowledge around technical debt, and support the drive to study this issue further.

In reviewing the body of research just described it is notable that a particular research gap exists. While studies have examined specific aspects of technical debt, including how technical debt accumulates during software development, decision factors for debt governance, and how technical debt can be characterized, previous work does not provide a systematic treatment of the characteristics of concrete technical debt instances. There is also very little work concerning the dynamics, the size and effects caused by technical debt when software development is ongoing. Similarly, while agile software development practices are said to be capable of withstanding (and mitigating) the effects of technical debt [23], previous work did not examine the effects of such practices on the occurrence and management of technical debt.

In fact, it remains unclear if and where technical debt is likely to accumulate the most in terms of the specific phases of software development. Ernst et al. [21] note that architectural components and early-stage decisions contribute to technical debt, however, whether these occur within a specific development phase, especially in relation to other software development phases, remains undetermined. As such, insights into these issues could be useful for those charged with managing technical debt. Beyond recommendations for practitioners, outcomes from such investigations would also enrich the knowledge pool and literature base around technical debt. We next consider some of the literature that has addressed agile software development, which sets the tone for our research agenda and the specific techniques that are used in the current work (refer to Section 3).

### 2.3. Agile software development

The Agile Manifesto [1] reflects a software development philosophy that emphasizes a context in which resources are scarce and requirements volatility is high. Given the many voices in support of agile methods, these approaches have become widely adopted and studied [24]. Agile development is implemented by several methods, and according to Dybå et al. [25], of the many flavors of agile, Extreme Programming (XP) [26] and Scrum [23] are two of the most widely studied. Equally, these two approaches are also considered to be the most frequently adopted in the software development industry [27,28]. We thus now briefly examine these approaches.

The XP method is seen to implement the agile manifesto's recommendations primarily through 12 practices, which are applied throughout software development. For example, the On-Site customer practice (see the Questionnaire http://soft.utu.fi/tds16/questionnaire.pdf , Q14 for the practices of the Extreme Programming method) calls for the customer to be always available. This practice is said to shorten feedback time, as developers can query the customer's opinion and resolve issues rapidly, resulting in fewer costly readjustments [26]. As a complement, the Scrum method defines processes and process artifacts. Here, customer feedback (as mentioned for XP) is implemented primarily in the *Iteration Review process* (see the Questionnaire http://soft.utu.fi/tds16/questionnaire.pdf , Q15 for the abstracted process components of the Scrum method), which calls for concluding each development iteration with a meeting wherein the *Product Owner* serves as a customer representative and provides feedback [23]. As Scrum defines processes and XP concentrates on practices, together these methods provide an adequate representation of agile software development [29]. We have thus used these methods as a basis for investigating specific software development practices and processes in this work.

## 3. RESEARCH APPROACH

In the following three subsections (Sections 3.1, 3.2, and 3.3) we explain our research approach. We first define our research questions in Section 3.1, and describe how the study is designed in Section 3.2. Thereafter, in the final section (Section 3.3), we describe how and why the study is conducted across three countries (Brazil, Finland, and New Zealand).

### 3.1. Research questions

Previous reports [20,23] indicate that the "technical debt" metaphor has been applied by some practitioners in the software development industry, and that this metaphor may also be readily understood by non-technical team members (e.g., customers, managers, and sales personnel) [21]. In fact, it has been contended that such metaphors could help to close the communication gap between technical and non-technical individuals and teams (including project managers) [7]. The usefulness of such knowledge would of course depend on individuals' perceptions about technical debt, which may be influenced by their backgrounds. However, to the best of our knowledge, previous work has not considered this issue, especially from a role- and background- specific perspective.

We thus examine whether technical and non-technical stakeholders share the same understanding of the 'technical debt' metaphor, how the respondents have been exposed to it, and whether they perceive the metaphor to be useful, by answering the first research question:

> **RQ1** (a) Are there differences in various stakeholders' perceptions of technical debt, and (b) do stakeholders' backgrounds affect such perceptions?

Previous studies reviewed above have focused on particular software development contexts and made general observations about technical debt in such settings [14,15,19,20]. In one instance the views of a few specific practitioners were captured [14], while in another change control boards were engaged [15]—both studies thus provide limited access to practitioners' opinions. While others have sought input from a larger spread of practitioners [20,21], these members were also drawn from homogeneous contexts (i.e., single or few companies, based on company sizes, or development areas). In terms of growing our general understanding of technical debt, it is pertinent to ascertain how a larger cross-section of the software development community perceives technical debt in their day-to-day work. In particular, we currently lack a comprehensive picture of technical debt's drivers (e.g., the reasons for its emergence as well as the influence of continued software development) and the software development phases it affects. If a broad understanding is to be pursued in these matters then researchers should strive to examine a range of perspectives, from those across company, sector, scale, and country boundaries. We thus outline our second research question to characterize technical debt and to address this issue:

> **RQ2** (a) What are the perceived drivers of technical debt, and (b) which software development phases does technical debt affect?

In spite of the growing predominance of agile methods in industry practice, previous studies have not reported practitioners' perceptions regarding the capability of agile methods in terms of the management of technical debt. As a method is composed of the practices and pro- cesses it introduces, there could be specific agile practices or processes that mean a development effort is able to withstand or mitigate the effects of technical debt. Further, the possibility of finding practices and processes that work towards the opposite end, increasing technical debt and/or amplifying its effects, is equally possible. For instance, the drive to deliver working software after each iteration may result in many shortcuts, leading to the accrual of technical debt. Validating this proposition could be useful for the community, in terms of exploiting the identified agile practices' and processes' strengths and weaknesses in the management of technical debt. We thus outline our third research question.

> **RQ3** (a) Are practices and processes of agile methods perceived to have an effect on technical debt or its management, and (b) which practices are deemed to have the most significant effect?

Among the insights that are likely to result from answering the research questions outlined here, understandings of practitioners' perceptions regarding technical debt could be useful in informing targeted education strategies. In the same vein, insights into the drivers and software development phases that contribute the most technical debt could sharpen developers' oversight towards reducing future instances of technical debt. Further, knowing which agile practices mitigate or amplify technical debt could provide useful pointers and inform developers' strategies in terms of which practices to use and when. We draw independent motivation for the examination of research questions such as these from Behutiye et al.'s [22] call to provide more solutions for managing technical debt in the agile software development context.

**3.2. Study design and background**

The following content describes the instrument that was used to gather the industry practitioner answers to our research questions as outlined in the previous section. Specifically, a questionnaire (see http://soft.utu.fi/tds16/questionnaire.pdf) was developed to enable collection of suitable data. Except for contact details, the questionnaire versions used in Brazil and New Zealand are identical. The Finnish version differs from these versions by having four one word differences (e.g., to the question of "How many concurrent projects do you have?" the Brazil and New Zealand versions have an answer option wording "None (only one)" whereas the Finnish version has the wording "One" instead). Additionally, one question from the Finnish version was split into two for the Brazil and New Zealand versions, but content coverage is identical across the two versions. These changes were made to streamline answering, but the modifications were not propagated to the Finnish version as it was frozen slightly prior in preparation for execution. Below, the next three subsections describe the different parts of our instrument, and how their design was related to our objectives.

**3.2.1. General information**

The first part of the survey solicits general demographic information from respondents, relating to the responding individual as well as the organization in which they work. On the individual level, the respondent is queried about their total experience with software development and the roles they would typically assume. Experience is captured as a single choice from three options: under three, three to six, or over six years, which provides a rough division into novice, intermediate, and experienced practitioners, following the classification used by Salo and Abrahamsson [30]. The role aspect covers common software project roles, and is adopted from the classification promoted by Bruegge and Dutoit [31]. Each respondent may indicate any combination of these roles, or specify new ones. The roles are intentionally described independent of specific development methodologies (e.g., Scrum defines a Scrum Master role [23] which is closest to the Facilitation role in our survey), as agile practitioners are perceived to adapt practices and processes as needed. Under such circumstances roles may be defined quite differently, and in contrast to pre-described methods. Additionally, capturing roles that are not directly related to specific agile methods enables us to better understand the array of approaches software organizations adopt.

At the organizational level, respondents are queried about their host company and the work they do. For the company, its size, number of concurrent projects, and concurrent projects per team are recorded. The respondents are then asked to focus on the development effort with which they are most closely affiliated. For this endeavor, the nature of the software product deliverable is established (i.e., whether it is complete or stand-alone, partial, or another type of product) together with the delivery target (i.e., an external or internal client), in addition to the desired software development characteristics (i.e., transparency, predictability, efficiency, sustainability, and agility). This set of characteristics appears three times in the survey to capture respondents perceptions in relation to varying preferences and priorities. The characteristics are deliberately introduced with no formal definition, on the assumption that the target group of skilled practitioners is aware of these principles. We were also cautious that lengthening the instrument would discourage participants' willingness to complete it in full. Finally, the respondents are queried about their team's size, as well as both release and iteration cycle times.

Subsequently, practitioners' perceptions of technical debt are mapped to their general information (both individual and organizational) in answering RQ1. Beyond answering RQ1, the data captured in the general information section are also used to explore country-specific differences.

**3.2.2. Development techniques used**

The second part of the survey questionnaire focuses on agile software development techniques, and is concerned with establishing which agile practices and processes are applied by respondents. As querying an exhaustive list of available development techniques is neither sensible nor feasible, due to concerns over respondent motivation [32] and the numerous synonyms and customized techniques used, the most commonly employed agile practices and processes are used as a basis for our questions. We established in Section 2.3 that XP practices together with Scrum processes (as depicted in Figs. 6a and 6b respectively) cover the underpinnings of agile software development well, in addition to being frequently used [27,28]. Hence, respondents are presented with these options. For each option, the adoption level is recorded on a five-point Likert scale, as recommended by Alreck and Settle [32], while the adoption level descriptions used are based on a relevant prior study [30].

After capturing the development practices and processes applied we next seek respondents' perceptions of their capability in meeting a team's software development and management needs. Finally, the questionnaire solicits details related to technical debt (considered next). This part is introduced last, as we are exploring the effects of individual practices and processes on technical debt, and we are hopeful of avoiding opinion re-adjustment [33] with respect to the capability levels queried earlier. The information collected here, and in the subsequent section, is used to answer RQ2 and RQ3.

### 3.2.3. Technical debt

The final section of our multi-part survey questionnaire comprises questions related to technical debt. This section is divided into two parts; the first seeks to establish the respondent's technical debt knowledge, while the second focuses on their recollections of a specific instance of technical debt and its effects.

Establishing each respondent's technical debt knowledge in the first sub-section is enabled by querying their existing knowledge on the matter, supplemented by an optional text area to capture each respondent's description of the technical debt concept. Further, respondents are asked in which media, and in which development scenarios, they have used the term or seen it being used. Having established existing knowledge, respondents are then asked to read two technical debt definitions (those of McConnell [7] and Brown et al. [10], as described in Section 2.1) and to indicate if their own description conforms to these definitions. (We noted that Ernst et al. [21] baselined their respondents with the former definition.)

The second subsection addresses respondents' recollections of concrete technical debt and its effects. The respondents are first asked if the software development efforts they work on are affected by technical debt. The above-mentioned agile practice and process lists (noted in Section 3.2.2) are reiterated, and respondents are asked to classify the effect that each practice or process has on technical debt. The effect is recorded as a choice from a five-point Likert scale ranging from very negative to very positive. The concept is explained to respondents via examples—a practice or a process has a positive effect on technical debt if it "can for example enhance technical debt management, lower its accumulation, or decrease its effects". The opposite definition is given for a negative effect.

The final part of the second subsection asks respondents to provide a description of a technical debt instance that has affected their work. Similarly to Ernst et al. [21], we ask the respondent to limit their consideration to a particular software development setting. For the technical debt instance, the development phases in which the instance resides are queried, following the phase classification of Bruegge and Dutoit [31] comprising Requirements elicitation and analysis, Design, Implementation, and Testing. Further, the causes underlying the prevalence of technical debt are captured as a set selection. To facilitate this selection we have converted the element list provided by Kruchten et al. in the The Technical Debt Landscape (see Section 2.1) [8], and we have adapted it into a set of causes; interpreted as inducers of technical debt in the landscape's elements. It should also be noted that we have included those associated areas that the landscape discusses but did not classify fully as being technical debt. This choice is made in order to be able to record if practitioners' perceptions are in line with Kruchten et al.'s delimitation into visible and mostly invisible causes.

Respondents are also asked if the instance resides in a component that is considered to be either internal or external legacy. Finally, the dynamics of the technical debt instance are investigated by querying if continued software development affects its size and/or the magnitude of its effects. In querying the instance's dynamics, the size is left intentionally vague, and is situated immediately prior to the question regarding the effects' magnitude so as to draw attention to their separation. Our reason for taking this approach relates to the argument that technical debt may be intentionally allowed to grow in size, and its effects may only be felt at the time of realization [34].

While a combination of the answers to the questions outlined above is used to tease out answers for RQ1 and RQ3, the feedback captured via the questions in this section is used in answering RQ2.

### 3.3. Study implementation and data collection

Our study was conducted through a web-based questionnaire. We deliberately chose this channel so as to minimize data transcription errors, while maximizing usability for the respondents [35]. Google Forms (see: https://www.google.com/forms) was chosen to host the survey and to enable data collection and pre-processing.

The questionnaire contained 37 questions, 35 of which are closed (refer to http://soft.utu.fi/tds16/questionnaire.pdf for a copy of the survey). The respondents could choose to define a concrete instance of technical debt by completing the optional part of the survey, but otherwise, an answer was required for all closed questions posed. The open-ended questions were used to prompt respondents for further details. However, very few respondents answered these questions, as we note below.

From the above description it is notable that our survey deviated somewhat from previous surveys (as per the studies considered in Section 2.2), where more open-ended questions were used to solicit practitioners' feedback. Such an approach was used to enable us to obtain a more definitive identification of technical debt effects for the pre-defined agile practices and processes, in addition to capturing information about concrete instances of technical debt in a more structured manner. That said, most questions included an open, "other", option so that respondents could provide all forms of answers.

After survey construction, the questionnaire was trialled within the authors' organizations (one trial in each country: Brazil, Finland, and New Zealand). Adjustments to phrasing and answer options were made to ensure consistency and clarity in terms of country-specific interpretation of questions and answer options. While only one of the target countries had English as its first language (i.e., New Zealand), trials across all territories with a single English version resulted in a low number of (mis)interpretation errors, as discussed in the next paragraph for Finland and Brazil. As such, the English version was used for all territories to enhance consistency.

A two-part pilot test was undertaken with software practitioners in Finland. The first part of the pilot was undertaken as an on-the-spot survey, with the Finland authors recording possible issues the respondents had with the questionnaire. The second component of the pilot was to execute the survey in a software incubator wherein multiple independent start-ups completed the questionnaire. Again researchers were present to record possible issues. These two pilot phases, both with the same instrument, gathered twenty answers. The misinterpretation error rate was calculated as the portion of questions having any sort of issues brought up divided by the total number of questions. For Finland, this figure was under 3%. Surveyors in Brazil discussed having observed misinterpretations similarly, but no exact results were recorded for their pilot. Finally, based on the pilot answers, we confirmed that the pilot respondents had varying backgrounds in terms of years of experience, as well as their current use of conventional and agile methodologies. As we wanted to probe technical debt opinions from experts working with varying software development methods and having varying levels of software related expertise, it seems that the pilot matched the target population rather well. The responses reported for the conduct of the full survey closely reflect those obtained in the pilot, indicating a good degree of comparability between these sets.

For each country (Brazil, Finland, and New Zealand), a questionnaire service was set up. Access to this service was anonymous for respondents. The respondents' participation was solicited from industry-affiliated member groups, mailing lists, magazines, and research partners (e.g., direct contact to the Finnish software entrepreneurs guild members and flyers in the Agile Day Conference series and meetings of the Software Measurement Association events). A cover letter was sent to all participating organizations, explaining the objectives of the study, where the study collaborators were introduced. A privacy policy was also presented to all respondents at the beginning of the questionnaire. Respondents from all software development backgrounds were welcomed to take part in the survey. Later sections of the paper acknowledge, for example, non-agile environments by filtering the dataset according to the respondents' backgrounds.

Respondents were allowed up to three months to complete the survey (through to May 2015), and data analysis began after the survey was taken offline. All data from the three countries were consolidated into a CSV file. Answer coding was applied where required (e.g., transforming Likert scale entries to integers in order to allow for ordinal comparisons) prior to combining and analyzing the respondents' entries in order to derive answers to the previously posed research questions. Our results and analysis are presented in the following Section 4.

## 4. RESULTS AND ANALYSIS

As per our study design we collated the three datasets: from Brazil, Finland, and New Zealand. The Brazil dataset comprised 62 completed questionnaires ($N_{BRA} = 62$), Finland's had $N_{FIN} = 54$, and New Zealand's $N_{NZL} = 68$. Based on the target group analysis (considering the number of individual email addresses used and company size estimates for non-direct emails) we estimate the response rate to be around 15.8% for Finland and 13.6% for New Zealand. A more obfuscated target frame unfortunately disallowed producing this figure for the Brazilian data set (i.e., we could not attain readership figures for the forums in which the survey was advertised). Due to the unsolicited nature of the survey invitation, the response rate is rather low—as indicated to be the case for other similar surveys [36]. However, on the positive side the use of an online tool allows for targeting of a much larger sample (i.e., targeting industry-affiliated member groups and mailing lists) which yields—in comparison to other surveys reviewed [14,16,18]—a much greater total response count ($N = 184$ for this survey). Further, all gathered questionnaires were complete, and so we did not need us to delete any responses from the set.

We analyze these responses to answer the research questions posed in Section 3.1. General, characterizing observations are presented in Section 4.1 based on the country-wise datasets and the collated dataset. The three research questions are then answered based on the collated dataset in Sections 4.2 through 4.4 respectively.

### 4.1. Background characteristics for the datasets

For all of the following results we provide a summary of the relevant statistics in our complementary online files (see http://soft.utu.fi/ tds16/backgrounds.pdf) to enable further review. To contextualize our results we first establish the magnitude of the respondents' software development activities by querying the size and the number of concurrent projects undertaken in their organizations. Very few of the respondents indicated having only *one* project (circa 5%) while *2 - 10* concurrent projects was the mode (circa 50%). Larger number of con- current projects was also quite common (*11 - 25* projects for circa 15%, *over 25* projects circa 20%).

The distinct country-specific distributions for the number of employees in organizations were also queried to further probe the noted pattern: the Brazil distribution has two peaks in the 10 - 50 and over 250 employees categories (circa 30% and 40% shares respectively). The three categories *under 10, 10 - 50*, and *51 - 100* employees evenly capture circa 90% of the Finnish respondents' organizations, while half of the New Zealand responses indicate an organization size of *over* 250 employees (smaller size categories capture the other half quite evenly). Of note is that a corresponding country-specific significant statistical difference was not observed.

The number of concurrent projects undertaken by teams and the size of respondents' teams were queried next. We note that all teams typically worked on *two to five* concurrent projects (circa 60% in this category, circa 26% in the only *one project*, and remainder in the *6 - 10* and *over 10* project categories evenly). For Brazil and Finland the average team size is smaller (both have a circa 45% peak in the *2 - 5* members category) than those reported for New Zealand (similarly proportioned peak in the *6 - 10* members category). The distributions are otherwise similar between countries. The majority of recorded team sizes are between two and ten persons. In their study, Ernst et al. found this category to be the second most common while the 10 to 20 people category was the most common team size. Examining the results for development iteration length, Finland and New Zealand seem quite similar, with an average iteration length of *2 - 3* weeks (circa 58% share), whereas Brazil's outcome (while also having a 26% mode in *2 - 3* weeks) is more variable across the length categories (*1, 2 - 3, 4, 8, over 8, no iteration* option range). Lesser differences are evident when comparing the average project lengths: all countries demonstrate a rather even distribution between the *1 - 3, 4 - 6, and over 6 months* categories (circa 30% share in each). From this, projects captured for Finland seem to be slightly shorter (peaking at *1 - 3 months* with a 37% share) while Brazil's are longer (peaking at *over 6 months* with a 37%) and New Zealand resides in the middle of the two. We suspected that organization size might be an explaining factor here, but no significant correlation was found in statistical analysis.

Further, software delivery arrangements also seem to be related to company size, as we examine the common project deliverable and its target in this light. Tending towards smaller companies, the responses from Finland demonstrate a greater emphasis on delivering complete software products (over 80% share) to external clients (over 60% share) in comparison to both sets of respondents from Brazil and New Zealand (where for both countries *complete software products* have a 60% share, and 40% of the delivery is done for an external client).

Examining the respondents' backgrounds, we note the following when querying respondents' years of experience working with software development related activities: country-wise deviation is almost non-existent here and almost two thirds of respondents had *over six years of experience* (*3 - 6 years* following with an average 14% share, and *under 3 years* with a 9% share). This category places most of our respondents in the 'experienced' category as per Section 3.2.1. Ernst et al. [21] similarly found their respondents to have on average over six years of experience.

Wide distribution of results was observed for the number of employees working in respondents' organizations, and our respondents' average project duration indicates that the survey covered a broad organizational spectrum. Convergent distributions were also present in respondents' background details, for example, in years of experience. While this could be interpreted as being somewhat indicative of the global state of experience present in organizations, it should also be noted that 75% of the responses have been contributed by software practitioners who have been active, and hence, influenced by organizational cultures, for at least six years.

### 4.2. RQ1: technical debt perceptions

In proposing RQ1 we are interested in establishing if there are differences in stakeholder perceptions of technical debt, and in understanding if stakeholders' backgrounds affect those perceptions. The following subsections provide a two-part answer to this research question. The first focuses on establishing stakeholders' existing technical debt knowledge and its closeness to the definitions given while considering the respondents' backgrounds. The second identifies the communication media through which stakeholders have been exposed to the concept of 'technical debt', if they have applied the concept in various scenarios, and whether they find the concept useful.

#### 4.2.1. Differences in stakeholders' perceptions of technical debt

To form a baseline, the level of existing technical debt knowledge was queried from respondents using a five point Likert scale. Following this, a selection of technical debt definitions was displayed to the respondents with a similar Likert scale, so that they could indicate how close they perceived their definition of the concept of 'technical debt' to be to those presented. A Spearman's rank test indicated a significant correlation ($\rho = 0.389$, $p < 0.001$) between respondents' assumed previous knowledge and their conformity to the definitions shown, though the results here are moderate. This meant that many practitioners who noted that they have extensive knowledge about technical debt, also indicated that their perception of this subject closely conforms to the definitions shown.

In drawing on the respondents' backgrounds, we found that consideration of their assumed software development role led to differences in results: the definition closeness of those occupying *Client representation, Facilitation,* and *Other* roles deviated significantly from the total population's distribution ($\chi^2$ ($df = 4$) ≈ 18.92, 16.82, 152.02 for the aforementioned roles respectively). A general overview of the results, with respondents' existing knowledge and perceived closeness to the given definitions, is depicted in Fig. 1a. To enable role-wise analysis Fig. 1b displays the role distributions, while the knowledge subdivisions are presented in Figs. 2a and 2b.

Concentrating on prior technical debt knowledge first, we can observe a rather even distribution of attitudes in Fig. 1a. Categories well and adequately knowledgeable have recorded the highest responses, over 20% share. However, a rather high proportion of respondents, circa 15%, indicate that they possess no *[existing technical debt] knowledge*.

When investigating conformity to established technical debt definitions (Fig. 1a), we note that both extremes attract the respondents' answers: the categories of respondents' indicating their closeness to the given definitions to be *far or very far* both accumulate less than a 10% share of the answers. By comparison, both extremes of very close and close as well as *no knowledge* attract more answers, with close to 80% falling across the very close and close categories. It should be noted that the technical debt definitions have been shown to the respondent prior to this question. This may help the respondent's justification for choosing one of the extreme options, as he or she has an explicit definition to reflect his or her opinion against.

We apply the role-based subdivision to the previously presented data and present the results in Figs. 2a and 2b. For these radar charts, the two highest and the two lowest answer categories have been combined so as to not clutter the charts. Looking first at existing knowledge in Fig. 2a, it shows that respondents who associate with the *Design, Client Representation,* and *Facilitation* roles perceive to have the highest knowledge; however, these results were within a very small margin to the others. The lowest number of no knowledge answers is indicated by those respondents in the *Design and Management* roles.

Role-based subdivision of indicated closeness to given definitions is given in Fig. 2b. No major differences exist between roles; as was the case in the previous result. The *Management* role seems to slightly excel the others having

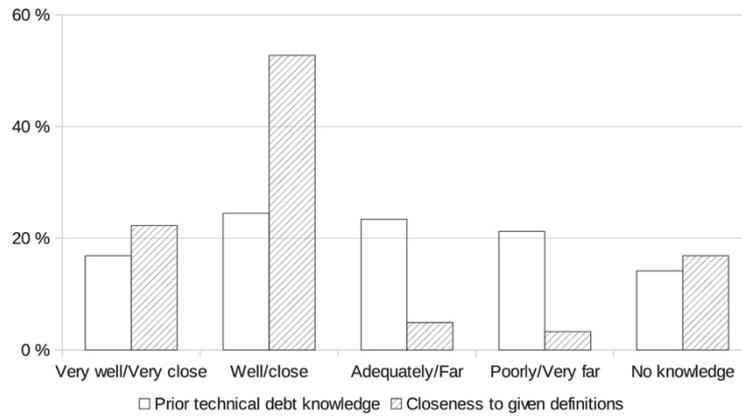

(a) Respondents' existing knowledge of technical debt and perceived conformity to presented technical debt definitions

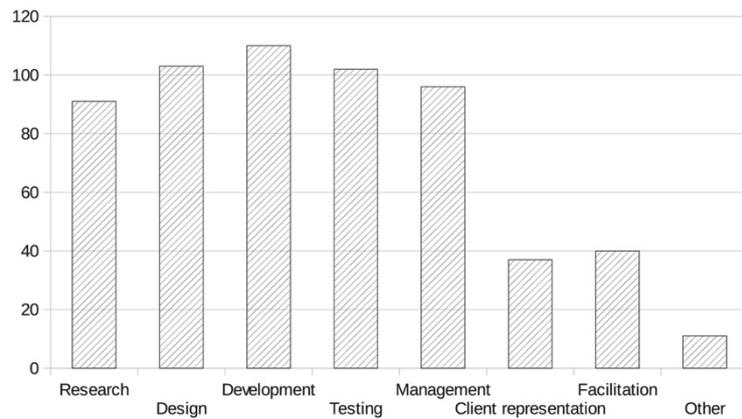

(b) Distribution of respondents' role association. Average number of roles assumed per respondent ≈ 3.21

Fig. 1. Background details for respondents.

the highest count in the close or very close category and at the same time the lowest count in the no prior knowledge category. In examining closeness to given definitions, it should be noted that these results only indicate respondents' proximity to those features of technical debt that the particular definitions captured (discussed in Section 2.1).

In formally evaluating our visual analysis we conducted Spearman's rank correlation tests, relating the respondents' backgrounds to their technical debt knowledge. For assumed software development roles and conformity to given technical debt definitions, respondents associating with the *Management* role indicated closest conformity of all roles (Spearman's rho; $\rho = 0.20$, $p = 0.007$). Further, more experienced respondents reported both better existing technical debt knowledge ($\rho = 0.18$, $p = 0.013$) and closer conformity to the given definitions ($\rho = 0.18$, $p = 0.017$). Lastly, regarding existing knowledge on technical debt, respondents working in larger teams ($\rho = 0.21$, $p = 0.005$) and developing complete software systems ($\rho = 0.15$, $p = 0.038$) reported greater existing knowledge than others.

**4.2.2. Communicating the concept of technical debt**

In order to understand how respondents have been exposed to the 'technical debt' concept, we establish the channels through which they have encountered this term (Fig. 3a), on whose initiative, and for which reasons (Fig. 3b).

Fig. 3a reveals that over 40% of the respondents have either seen or heard the term used in Work related situations. However, Field-specific or scientific literature eclipses these with over 45% share. Both accounts are interesting as they indicate a solid footnoting for the concept in the context of contemporary software development.

News media is the least popular choice with circa 13% share of the responses. Observing an increase in this share in the future would likely indicate that the concept's use in more general contexts and for a general audience has increased. The share of respondents who have not seen the term used is also quite high, totaling 18% of the respondents.

In Fig. 3b, for all provided software development related situations, 30% to 40% of the respondents indicate that the technical debt concept was used by them as well as by their colleagues. Notably, however, a circa 38% portion also indicates that the concept has not been used. In comparison, perceived usefulness is almost 25% higher than the level of current application, for all cases. This seems to indicate that the current level of concept application is lower than what the respondents perceive to be its potential. Finally, there is also a 15% share of respondents who perceive there to be no gains to be had from using the 'technical debt' concept.

**4.3. RQ2: technical debt drivers and affected software development phases**

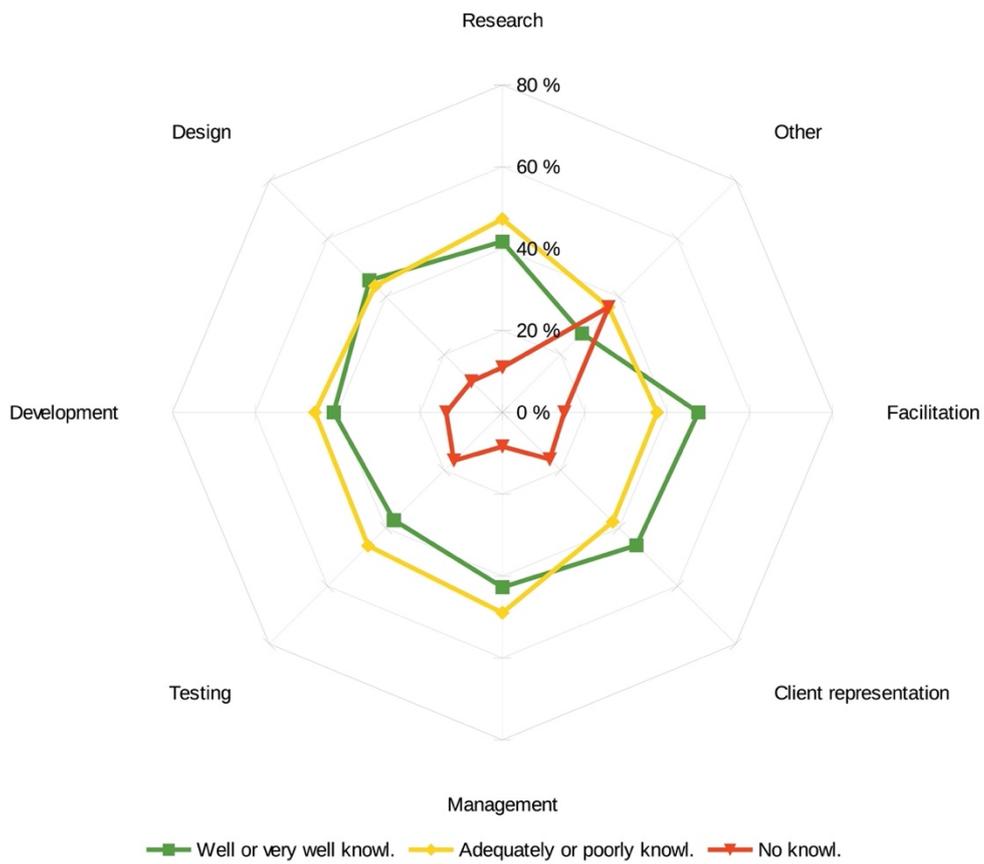

(a) Respondents' indicated existing knowledge of technical debt as a function of assumed role

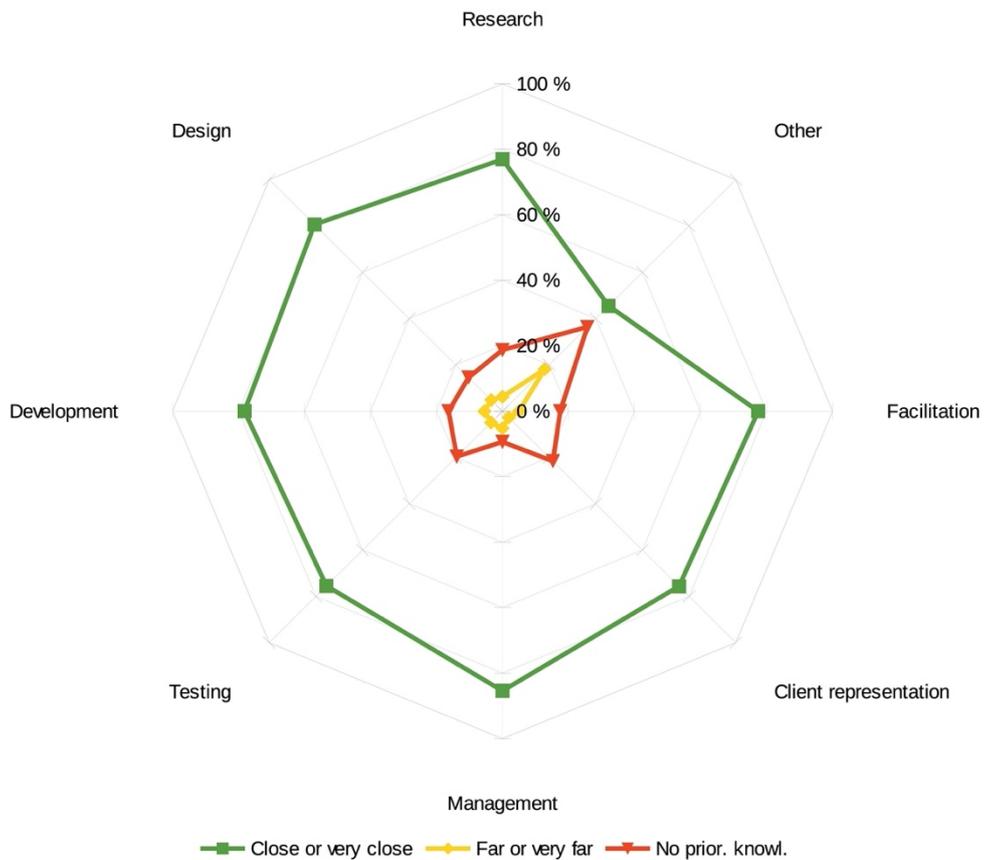

(b) Respondents' claimed conformity to presented technical debt definitions as a function of assumed role

Fig. 2. Role-based opinion distributions.

The second research question addresses the perceived drivers (causes and software development dynamics) of technical debt, and in which software development phases technical debt is seen to be more prominent. This is done by identifying and characterizing concrete instances of technical debt encountered by respondents in their software development endeavours. As per the study design, a description for a single technical debt instance was queried. From the Brazil respondents 22 ($N_{iBRA}$ = 22) could provide an instance description; circa 35% of the country's total responses. The corresponding figures are $N_{iFIN}$ = 16 (or c. 30%) for Finland and $N_{iNZL}$ = 31 (or c. 46%) for New Zealand. Noteworthy is that overall only around one-third of the respondents provided the queried description. While for many respondents a limited amount of time to respond would likely be a prominent explanation, argument could be made for other causes as well, namely the possible difficulty in identifying a technical debt instance to describe. For the following subsections, the $N_i$ = 69 descriptions (i.e., $N_{iBRA} + N_{iFIN} + N_{iNZL}$, c. 38% from all answers) form the dataset referenced. This is used to identify the drivers and affected software development phases for technical debt.

### 4.3.1. Drivers and phases

Fig. 4 reveals the space of possible causes for technical debt's emergence, as queried in relation to the adopted Technical Debt Landscape (refer to Sections 2.1 and 3.2.3). In considering all responses, between four and five causes were typically cited as contributing to the emergence of a single technical debt instance. This is a notably high number, when taking into account that the space included nine options in addition to a free description, and that the causes span different areas of the software life-cycle. Cause-by-cause the dataset indicates highest rates for the *Architecture is inadequate* and the *Structure is inadequate* categories. To clarify, the respondents were asked to limit their responses to a system or system component affected by technical debt. Indicating architectural inadequacy as a cause means that technical debt was accumulated due to the target component having a non-optimal higher level design or design compliance. Similarly, the *Structure is inadequate* category applies to instances at the component structure design and design compliance level, generally considered to be a step lower than architectural design.

The systematic review provided by Behutiye et al. [22] also notes *"Architecture and design issues"* as the most frequently discussed cause of incurring technical debt in an agile software development context. These findings seem to support one another and indicate that the research as conducted, at least in the agile domain, reflects industry attitudes and needs.

The least frequently indicated causes of technical debt emergence are *Additional features are required* and *New features are required*. These causes are also those which the landscape's original authors noted had limited contribution to technical debt; indicative of the landscape's success in delimiting technical debt causes from others.

In examining Fig. 5a, the most common origin of technical debt is indicated to be *Legacy from an earlier team/individual working on the same project/product*, with this category accounting for over 50% of the technical debt instances' origins. Around one quarter of the technical debt instances are seen not to originate from legacy components, and origins in other forms of legacy are also less frequent. However, we note a threat to validity here: these figures may be exaggerated due to respondents being more prone to selecting components that have foreign origins, and these components are often considered legacy (discussed further in Section 5.3).

As shown in Fig. 5b, from the queried software development phases, in almost 90% of the cases a technical debt instance is perceived to affect the *Implementation*, and in 60% of the cases the *Design*. The design can generally be considered emergent to the implementation. Not directly visible from the figure, a single technical debt instance was seen to span between 2 and 3 software development phases in general.

### 4.3.2. Dynamics of technical debt

In considering the views of the respondents, for circa 70% of the technical debt instances, continued software development is seen to induce an increase or a large increase in the size of the technical debt (Fig. 5c). Martini et al. [17] also establish similar dynamics as well as constant growth inducers for ATD. However, a portion of the respondents also indicated that circa 20% of the instances experience a *decrease* or a *large decrease* in size. It is possible that in these development environments, refactoring and other technical debt management procedures are continuously executed [17,21]. (We discussed the distinction between technical debt's size and the magnitude of its effect in Section 3.2.3.)

As can be seen in Fig. 5d, respondents generally tend to see that the effects induced by the technical debt instance are associated with its perceived size. A 40% share of the respondents indicate the size and effects to be directly proportional, while an almost identical portion of respondents perceive that a relationship exists but did not indicate the manner in which the technical debt instance's size alters the magnitude of its effects.

Finally, the respondents were queried if they perceived technical debt to affect certain characteristics of their software projects, comprising *predictability, efficiency, sustainability, transparency* and *agility*. A five point Likert scale was used to capture respondent opinions ranging from significant deterioration to significant improvement for the five characteristics. We performed data transformation by using an integer interval scale ranging from one to five in coding responses. In this case it was appropriate to analyze the full dataset ($N$ = 184), as this question was independent of respondents' ability to define a concrete instance of technical debt. Respondents generally perceived technical debt to have a somewhat deteriorating effect on all five queried software project characteristics (where the average answer fell between the options *Deteriorating effect(2)* and *No change(3)*, having a standard deviation of circa one option interval).

A $\chi^2$ test was then performed on the responses regarding the effect of technical debt on software development characteristics, and the results confirmed statistically significant relationships ($\chi^2$ = 63.32, $df$ = 20, $p$ < 0.001) to

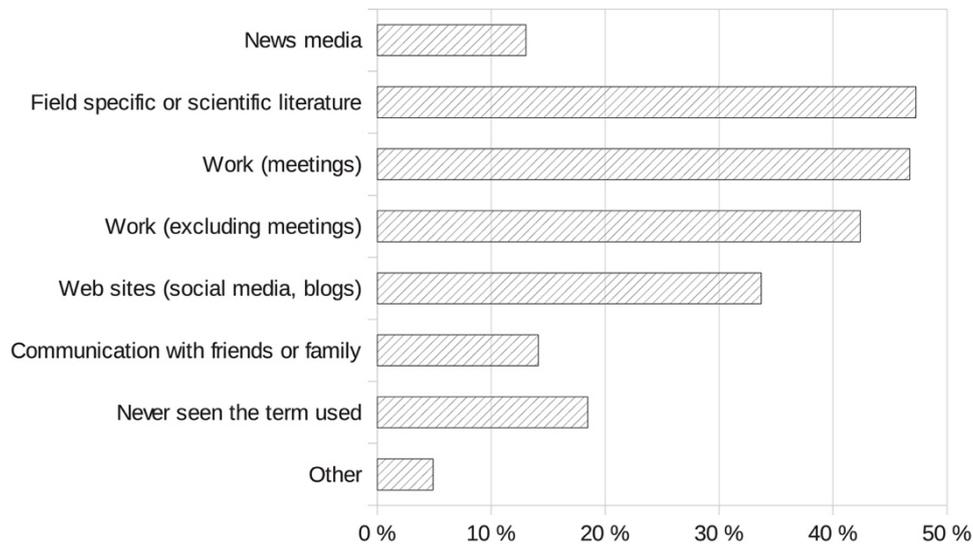

(a) Media in which respondent has seen or heard the technical debt term

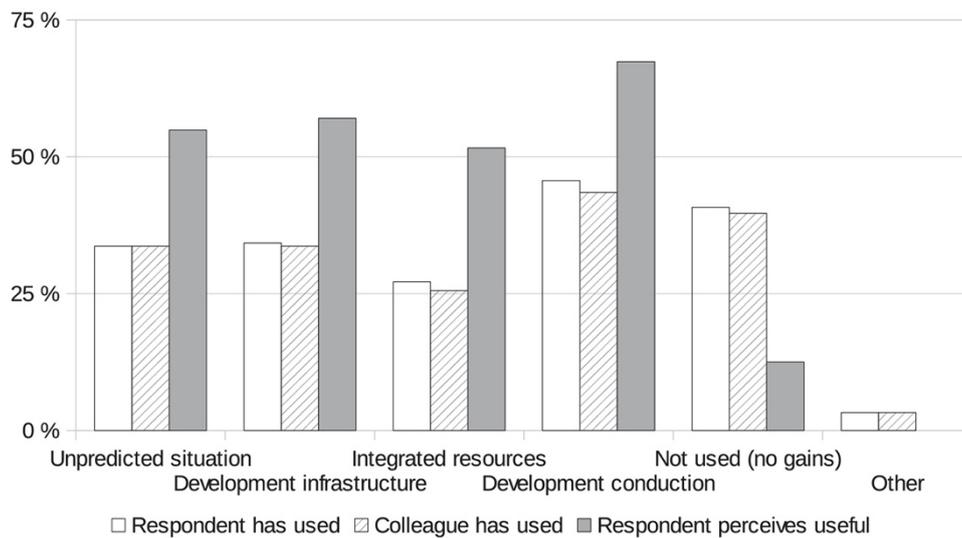

(b) Usage of the technical debt concept in software development related situations

Fig. 3. Technical debt usage.

exist between the *predictability, efficiency, sustainability,* and *agility* characteristics. Hence, even though technical debt is seen to have a deteriorating effect on all five queried characteristics, the effect on the *transparency* characteristic varies independently of the others. It should be noted, however, that the characteristics are provided in the survey instrument with no supporting definitions. As such, respondents may have varying definitions for the terms. This is discussed as a limitation in Section 5.3.

**4.4. RQ3: technical debt management and agile practices**

In addressing the final research question we are interested in establishing if agile practices and processes are perceived to have an effect on technical debt or its management, and if so, which are deemed to have the most substantial effect. To answer this question, we first establish a baseline for a specific set of agile techniques (the Scrum and XP techniques as discussed in Section 3.2.2), by surveying their level of adoption and their perceived ability to meet managerial requirements, and we then ascertain the techniques' perceived effects on technical debt. These issues are considered in the two subsections that follow.

**4.4.1. Perceived capabilities of agile practices**

Figs. 6a and 6b depict adoption levels for agile practices and processes respectively. *Coding standards, Continuous integration, 40-hour week,* and *Open office space* are the most frequently adopted practices, while Planning game, Pair-programming, and On-site customer are the least frequently adopted.

For the queried agile processes and their artifacts no notable differences are present. All queried items average an adoption level similar to the most adopted practices from the previous group. *Iterations* are used most frequently,

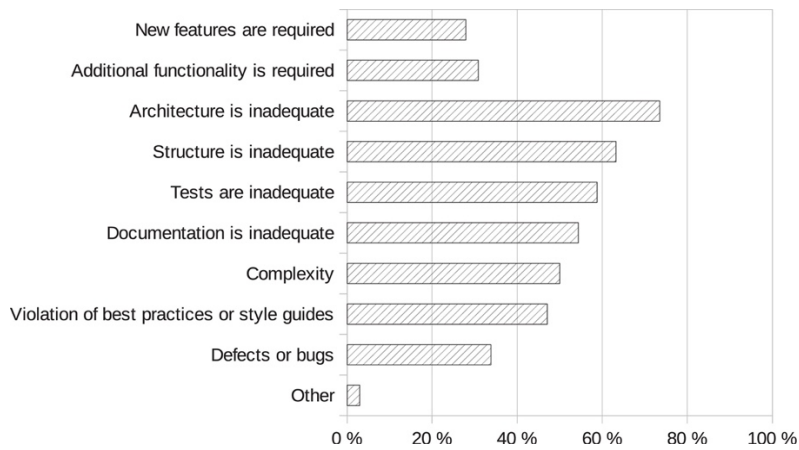

Fig. 4. Causes of concrete technical debt (based on a subset from total answers, $N_i$ = 69).

while *Iteration reviews/retrospectives* are less frequently adopted.

**4.4.2. Effects of agile practices and processes on technical debt**

In considering the impact of agile practices and processes, Figs. 7a and 7b respectively depict their perceived effects on technical debt. For both radar charts, the positive and very positive as well as the negative and very negative answer categories are combined so as to not clutter the charts. To ensure experience-based perceptions are reported, the results concentrate here only on data on adopted practices and processes (excluding answers where respondents have indicated that *Development (flow) practice is not used* in Question 36 & 37; see Section 4.4.1 for discussion on adoption).

Fig. 7a illustrates that, for many of the queried agile practices, the effect on technical debt is seen to be positive (c.f., Section 3.2.3). In particular, the practices of *Coding standards* and *Refactoring* are indicated as having a positive effect by over 75% of all respondents. Behutiye et al. [22] find refactoring and coding-related issues to be frequently discussed topics as both causes of technical debt (when there is a lack of application) and as prominent technical debt management strategies, at least in the agile domain. This indicates that research on the subject is inline with what practitioners perceive as being important for technical debt management.

A slightly less positive perception is observed for the *Continous Integration, Test Driven Development,* and *Collective code ownership* categories. A neutral effect is observed for the *40-hour week*, *Open office space*, and *Planning game* practices. Finally, the most negative effect is observed for the *On-site customer* practice. The negative share is less than 20%, but the highest dispersion in effect opinions is also recorded for this practice.

The effects of agile processes and process artifacts on technical debt are much more uniform in comparison to the practices, as is evident in Fig. 7b. Whilst a generally positive effect is indicated for all processes, the *Iteration reviews/retrospectives* technique enjoys the strongest positive and the lowest neutral opinion. This could be seen to support the finding of Ernst et al. [21] where 31% of respondents indicated retrospectives to be the point of identification for technical debt. The Product backlog process artifact seems to lead on negative effects but even in this case the share is less than 10%.

Table 1 reports, for the queried agile practices and processes, the statistically significant Spearman's rank correlations between each technique's perceived adoption level and its perceived non-negative or non-positive effect on technical debt. Both the adoption and the effect were recorded as selections from a five-point Likert scale (see Section 3.2.2). However, since the effect scale ranged from negative to positive effects, it was not directly comparable to the adoption scale. As such, the data was spliced into two for correlation testing: answers indicating non-negative effects and answers indicating non-positive effects. Finally, only answers that indicated some level of use for each technique were considered.

From Table 1 we note that statistically significant correlations ($p < 0.01$) are evident between the respondents' indicated adoption level and the non-negative technical debt effects of the *Collective code ownership* agile practice and the *Iteration reviews/retrospectives* agile processes. Correlation here indicates that a higher perceived adoption level for a technique is associated with perceiving more positive technical debt effects for the technique, or a lower adoption level is associated with perceiving no effect for the technique. In both cases, however, the correlation can be considered weak [37].

In Table 1, statistically significant correlations exist between the adoption level and the non-positive effects of the *Iteration planning meetings* and *Iterations* agile processes. This indicates that a lower perceived adoption level for a technique is associated with perceiving more negative technical debt effects for the technique, or a higher adoption level is associated with perceiving no effect for the technique. These correlations can be considered moderate [37].

At this point it is important to note that respondents' backgrounds could be considered a threat to validity. Respondents' existing technical debt knowledge can be interpreted as indicative of their ability to acknowledge and govern technical debt within their projects. This could have an effect on projects' technical debt, independent of the applied set of development practices and processes. Thus, the effect of agile practices and processes must be observed to be independent from the respondents' existing technical debt knowledge.

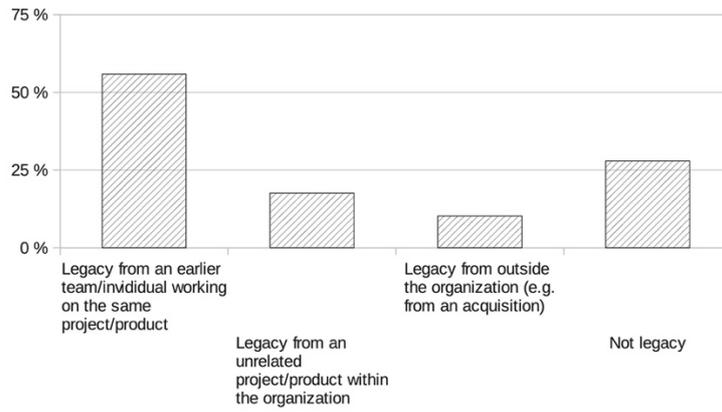

(a) Origins of concrete technical debt

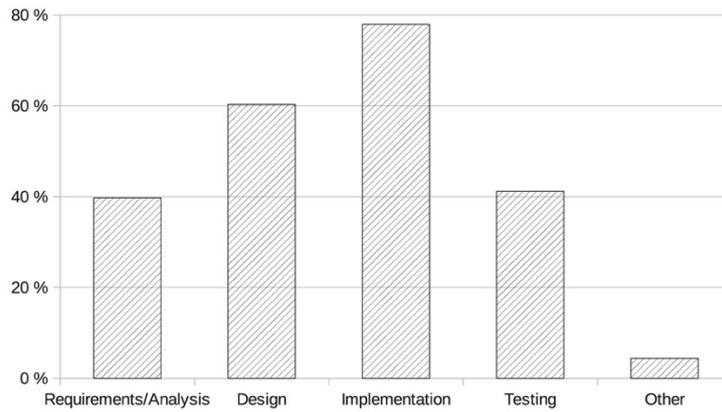

(b) Software development phases affected by concrete technical debt instances (average instance spans 2 to 3 phases)

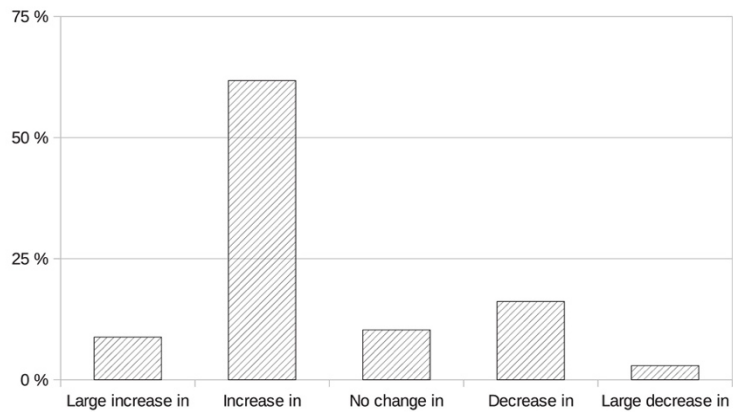

(c) Effect of continued software development on the size of concrete technical debt instances

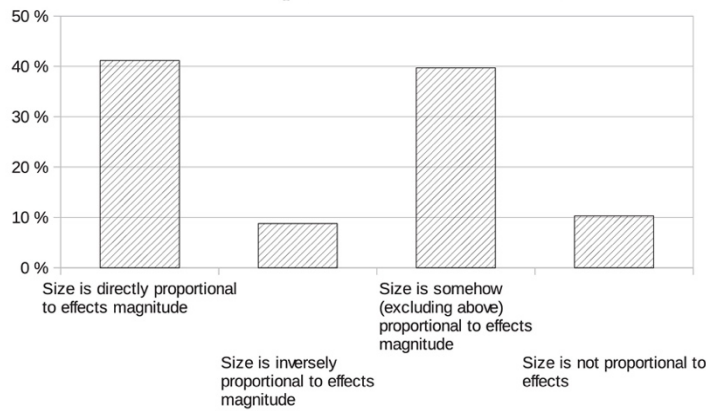

(d) Relationship between concrete technical debt instances' sizes and the effects they induce

Fig. 5. Characteristics for concrete technical debt (based on a subset from total answers, $N_i$ = 69).

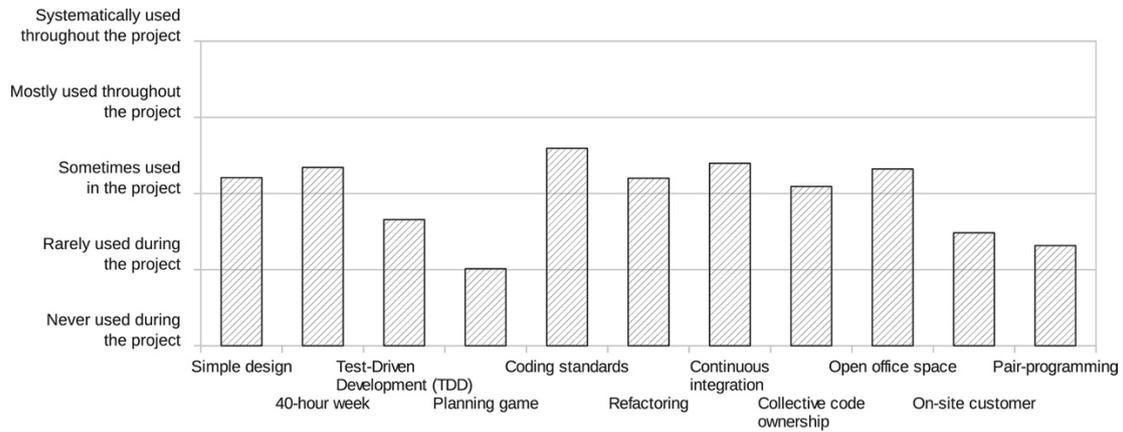

(a) For queried agile software development practices

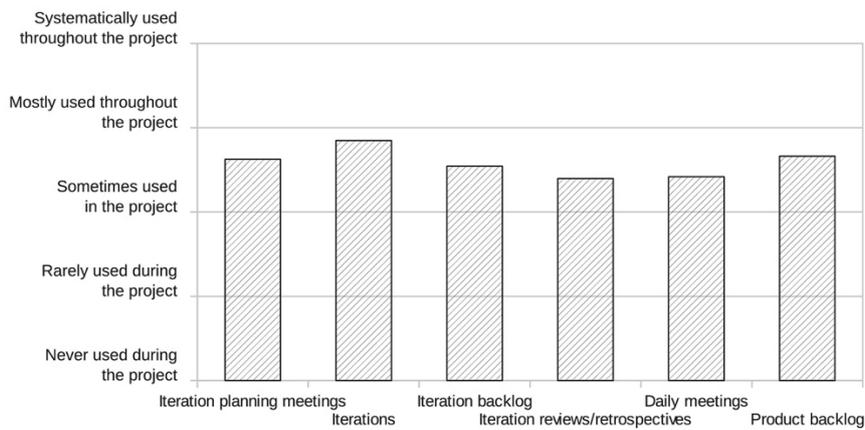

(b) For queried agile software development processes and process artifacts

Fig. 6. Average adoption levels for queried agile practices and processes.

Table 2 reports the statistically significant Spearman's rank correlations between respondents' existing technical debt knowledge (see Fig. 1a) and the perceived effect of agile practices and processes on technical debt (see Figs. 7a and 7b). As with Table 1 above, the data is spliced into perceived non-negative and non-positive effects prior to analysis, and only techniques that are in use are considered.

Table 2 lists those agile software development practices and processes for which statistically significant correlations (at the $p < 0.05$ level) are observed between respondents' existing technical debt knowledge and the agile techniques' indicated non-negative or non-positive technical debt effects. For the non-negative effects, the correlation conveys that a perceived higher technical debt knowledge is associated with the respondent indicating a more positive technical debt effect for the agile technique, or that a perceived lower technical debt knowledge is associated with the respondent indicating a neutral technical debt effect for the technique. The inversion applies for the non-positive correlations: perceived lower technical debt knowledge is associated with the respondent indicating a more negative technical debt effect for the agile technique, and a perceived higher technical debt knowledge is associated with the respondent indicating a neutral technical debt effect for the technique.

For all cases in Table 2 except for *Coding standards* and *Continuous integration* the correlation is weak [37]; the other correlations are moderate. The non-significant and weak correlations provide evidence for these agile practices and processes having a respondent-background-independent effect on technical debt. The moderate non-positive effect correlations are discussed in the threats to validity (see Section 5.3).

Spearman correlation coefficients were similarly calculated for the relationship between respondents' team size and the perceived effects of agile practices and processes. Only the *Planning game practice* (Spearman's $\rho = 0.192$; $p = 0.032$) and the *Iteration backlog* process artifact (Spearman's $\rho = 0.226$; $p = 0.005$) showed significant, but weak, correlation. Thus, queried agile practice and process effects seem to be independent of team size. We further discuss these and other findings in the following section.

## 5. DISCUSSION AND IMPLICATIONS

This section discusses the results of our exploratory survey, first highlighting our key findings in Section 5.1. Theoretical contributions and managerial implications are next reported in Section 5.2; this section also outlines

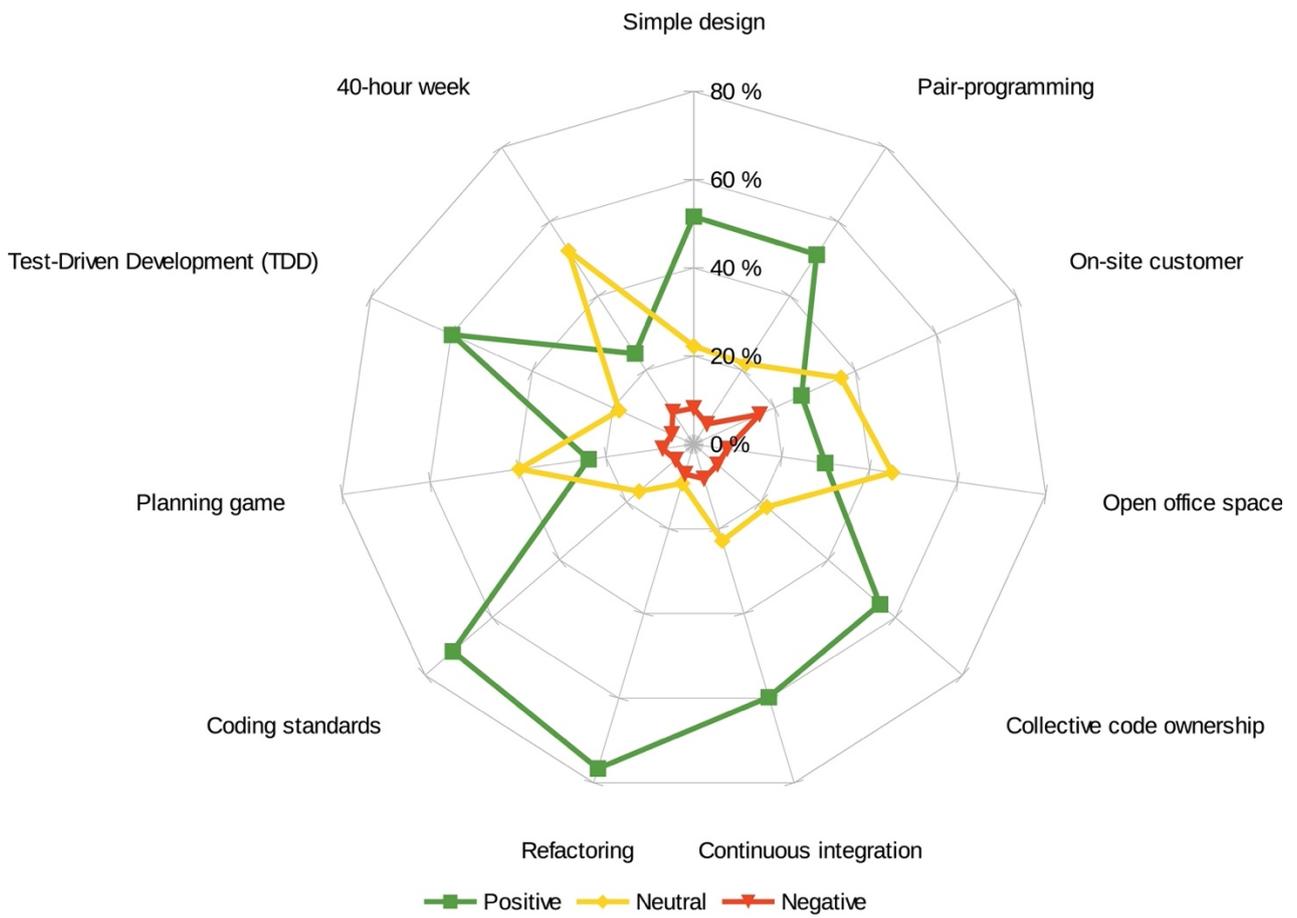

(a) For agile software development practices'

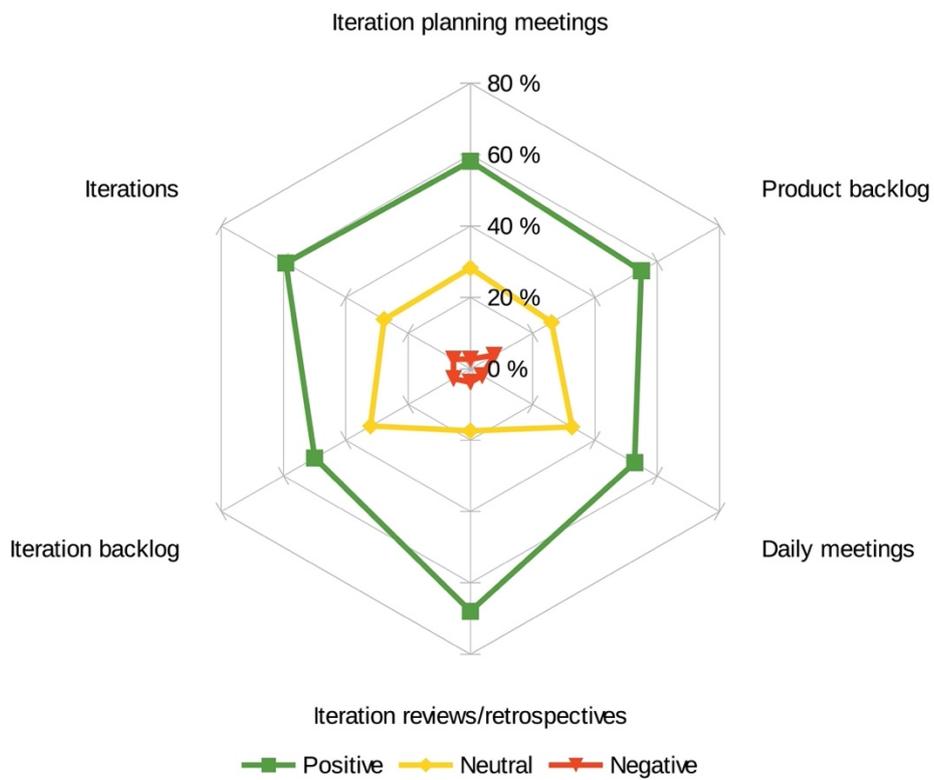

(b) For agile software development processes' and process artifacts'

Fig. 7. Perceived effects on technical debt.

avenues for future work. Section 5.3 then considers the limitations to the work.

**5.1. Key findings**

We outlined three research questions for this study in Section 3.1: RQ1 (a) are there differences in various stakeholders' perceptions of technical debt, and (b) do stakeholders' backgrounds affect such perceptions; RQ2 (a) what are the perceived drivers of technical debt, and (b) which software development phases does technical debt affect; and RQ3 (a) are practices and processes of agile methods perceived to have an effect on technical debt or its management, and (b) which practices are deemed to have the most significant effect? Effect is classified as withstanding, mitigating or amplifying in this work. Sections 4.2, 4.3, and 4.4, respectively, report our results in relation to the three research questions posed. In this subsection we revisit our key findings and subsequently discuss their implications for theory and practice.

**First key finding:** regardless of software practitioners' backgrounds we observe from their responses that they are largely aware of the concept of technical debt. We observed that while developers had moderate knowledge of technical, those that were aware of the concept appropriately defined same. The findings on developers' existing knowledge of technical debt deviate somewhat from those of Lim et al. [19], whose results indicated generally lower technical debt knowledge among their respondents. Our outcomes here suggest that there is increasing awareness of technical debt amongst practitioners. This increasing awareness around technical debt may be positive for practice in terms of the avoidance and management of events that may result in debt during software development. That said, awareness may not always translate to application, as we noticed in Fig. 3b that current application of the technical debt concept is considerably lower (circa 35% of respondents or their colleagues applied the concept) in comparison to its perceived usefulness (circa 55%).

It should also be noted that practitioners across roles had similar understanding of technical debt as a concept (see Fig. 2b), supporting the findings reported by Ernst et al. [21]. This evidence supports the premise that technical debt, at least to the extent of the commonly discussed definitions, is a concept that can facilitate software development-related discussions where all roles have mutual understanding (i.e., the concept of technical debt is understood consistently by all stakeholder groups). Hence, this could contribute towards reducing the communication gap between technical practitioners and other stakeholder groups. This potential benefit was also noted by McConnell [7].

**Second key finding:** technical debt is influenced by common agile software development practices and processes (i.e., practitioners perceive these techniques to have varying effects on technical debt; see Section 4.4.2). In particular, our evidence here indicates that agile practices and processes that verify and maintain the structure and clarity of artifacts (e.g., *Coding standards* and *Refactoring* practices) within software development projects are perceived to have a considerably positive effect on the management of technical debt. In fact, we observe such a finding for the majority of agile practices and processes in Figs. 7a and 7b.

We perceive that the influence of agile practices and processes observed is an important premise for advancing technical debt management. For the above identified agile practices and processes, practitioners' perceptions about technical debt need to be validated. An avenue for this is to closely track technical debt instances in environments that use these agile techniques, and to relate changes in the instances back to the techniques' use. For some techniques (e.g., the *Iteration backlog*), it is possible that the use of explicit or enhanced technical debt information (e.g., by entering technical debt instances as separate items into the backlog) could be studied as an independent variable in support of validation. Ernst et al. [21] reported findings which indicate that some technical debt management is already taking place in many of the existing processes, providing further support for choosing this validation procedure. Information produced by the validations can be used to argue for or against using particular techniques in particular software development contexts.

While the generally positive results speak for integrating agile software development practices and processes for technical debt management, areas of concern still remain within these techniques. We believe that the aforementioned communication gap and competing stakeholder interests (discussed in Section 2.2) are potential problem areas that may result in conflicts. More particularly, the most diverse, and in our case also the most negative, effects of technical debt management were reported for an agile practice that brings together multiple roles and responsibilities (*On-site customer* in Fig. 7a). This outcome supports the findings discussed earlier by Klinger et al. [14] regarding competing stakeholder interests. Competing stakeholder interests resulting in issues is interesting in light of the first key finding. Where there is potential friction between stakeholder groups here, there is mutual understanding between roles (i.e., possible stakeholder groups) in the first key finding. To pre-empt this contradiction, we argue, partially based on the findings of Martini & Bosch [16] on prioritization of different development aspects, that the stakeholder groups have differing motives and as such, while they may perceive technical debt similarly, they can operationalize the concept (e.g., value the technical debt) differently. Additionally, we did not observe a correlation between respondents' team size (which can contribute to the number of stakeholders present) and the perceived effects of technical debt. This suggests that practitioners may not need to make provisions for this factor when considering the management of technical debt.

There are undoubtedly also other aspects in addition to communication and competing stakeholders' interests through which we can analyse agile techniques' impact on technical debt. In particular, time pressure and requirements volatility are aspects that agile techniques try to combat against [1,23], but projects are still perceived to struggle with such issues. In fact, we observed that *On-site customer* and *40-hour week* practices invoked diverse opinions from respondents, while *Refactoring, Continuous*

Table 1. Statistically significant Spearman's rank correlations between respondents' perceived agile software development technique adoption levels and the techniques' (non-negative or non-positive) technical debt effects.

|  | ρ | p-value |  |
| --- | --- | --- | --- |
| Collective code ownership | 0.231 | 0.009 | non-negative |
| Iteration reviews/retrospectives | 0.221 | 0.007 |  |
| Iteration planning meetings | 0.366 | 0.009 | non-positive |
| Iterations | 0.415 | 0.001 |  |

Table 2. Statistically significant Spearman's rank correlations between respondents' perceived existing technical debt knowledge and the agile software development techniques' (non-negative or non-positive) effect on technical debt.

|  | ρ | p-value |  |
| --- | --- | --- | --- |
| Simple design | 0.185 | 0.031 | non-negative |
| Refactoring | 0.281 | $3 \cdot 10^{-4}$ |  |
| Iteration planning meetings | −.170 | 0.032 |  |
| Iteration backlog | −.165 | 0.043 |  |
| Daily meetings | −.176 | 0.027 |  |
| 40-hour week | 0.201 | 0.034 | non-positive |
| Coding standards | 0.407 | 0.009 |  |
| Continuous integration | 0.331 | 0.012 |  |
| Iteration backlog | 0.261 | 0.030 |  |
| Daily meetings | 0.269 | 0.028 |  |

*integration* and *Collective code ownership* practices were consistently reported to have a highly positive impact on technical debt management. On the other hand, the *40-hour week* practice can be problematic from a scheduling perspective as it tries to control overtime work, and the *On-site customer* can represent an unwanted time pressure element and it may increase requirements volatility. At the positive end, Refactoring can be seen to reduce schedule volatility if it reduces unforeseeable delays (as noted by Martini et al. [17]), *Continuous integration* minimizes lead time for completed features, and *Collective code ownership* removes bottlenecks as all stakeholder groups share responsibility and are capable of feature delivery during development.

**Third key finding:** industry practitioners perceive concrete technical debt to be difficult to identify, suggesting that it is complicated to manage. We observed that only a portion of the respondents provided a concrete technical debt description (subset of total answers where $N_i = 69$), indicating that technical debt instances are hard to quantify, and thus, manage. Technical debt instance seems to typically span more than two software development phases (Fig. 5b). In addition, a combination of several causes is responsible for technical debt emergence (Fig. 4). Furthermore, technical debt instances vary given the size of a software development project, however, in some cases the size is not directly proportional to the magnitude of the effect (Fig. 5c and 5d). In light of these findings, existing and new technical debt management procedures and tools should be scrutinized to evaluate if they are truly capable of considering all the identified characteristics of concrete technical debt. For example, is a commonly used management tool or procedure capable of accommodating technical debt instances that seem to reside in and affect several software development phases at the same time? Failure on the part of the tool could mean that particular aspects of the instance remain overlooked; making the management more cumbersome. A possibility even remains that unassociated stakeholders work on the same technical debt instance, unaware of shared technical debt across artifacts and phases. Further, the (sometimes) invisible nature of concrete technical debt may be part of the reason why technical debt instances are seen to grow in size during continued development. In fact, we note that most technical debt instances reside in the software implementation phase. As the implementation generally corresponds to the immediate value delivered to the client, growing technical debt presence can be seen to pose negative consequences on value delivery.

We should also strive towards validating the identified features of technical debt instances through other means. For example, for those technical debt instances limited to software components that have formal semantics available, the mining software repositories [38] methods can be used to track the propagation of the instance. Examining the propagation path and finding components from several software development phases would provide supporting evidence for our finding here, which observe technical debt instance to span over two software development phases.

Furthermore, for almost 75% of the technical debt instances considered by our respondents, their origins are in legacy software (Fig. 5a). While this figure may be biased due to inadequacy in practitioners' memory, the finding here is still of importance. Martini et al. [17] note that legacy is a notable contributor in the accumulation of architectural technical debt, and having established this level of interplay between technical debt and legacy in the results above, we note two implications for this relationship. First, legacy software should be researched as a potential interface for technical debt management. Software maintenance research may also lend support towards revealing mechanisms that identify, value, and govern legacy software. Second, the closeness of the 'technical debt' concept to the concept of 'software legacy' should be carefully examined to establish if practitioners fully distinguish between the two concepts. However, as software legacy is seen to be central to many technical debt instances, there is a possibility that the technical debt concept could be reduced to a synonym for legacy. To this end, the newer—possibly more popular—metaphor could be used to justify bad code whilst disregarding the investment aspects that are necessary for technical debt management.

### 5.2. Theoretical contributions and managerial implications

This study contributes to the ongoing academic discussion of technical debt in two respects. First, industry practitioners across multiple territories have shown high conformity (or can relate) to the given definitions for technical debt (i.e., the definitions of McConnell [7] and Brown et al. [10] queried for Fig. 1a). This evidence suggests that the technical debt concept is understood by both academic and industrial bodies. This observation provides an important premise regarding, for example, the applicability of academic solutions in industrial settings. Second, observing that respondents found the chosen selection of causes for technical debt emergence (i.e., the *The Technical Debt Landscape* queried for Fig. 4) to be suitable, both classification- and grouping-wise, for documenting instances of technical debt is indicative of its representativeness and potential utility. In addition, we observed the lowest selection frequencies for those causes that the landscape's authors have also excluded from being technical debt.

To the best of our knowledge, this study is the first to verify the work of Kruchten et al. [8], and provides support for its continued development and application. Further, the finding that a predefined selection of causes capture technical debt well is somewhat divergent from the previous findings of others (e.g., [20]) where practitioners felt that a self-produced taxonomy of technical debt worked the best.

Results presented in this work may also be of use to those governing the daily activities involved in software development and the management of technical debt. First, while we observed in our results that there are differences in the level of knowledge between personnel coming from differing backgrounds, uniformity in closeness to the given definitions and in the perceived usefulness of applying the technical debt concept gives a strong indication of the utility of the technical debt concept as a management tool applicable for inter-stakeholder-group communication and planning.

Second, our outcomes are likely to be useful to managers in terms of the effects that common agile practices and processes have on technical debt. While our second key finding discusses the effects of the practices in more detail, we note that the results provided herein should yield valuable information to project management when evaluating the applicability of different practices and processes when used in projects. As an example of this, we note that *Coding standards* was seen to have a very positive effect on technical debt management. In a recent survey [39], however, only 44% of respondents were using the practice in question. Thus, this study, among others, speaks to the importance of using coding standards; this time as one of the most important agile processes for mitigating the effects of technical debt.

We note that industry professionals, regardless of their role, should pay close attention to applying the technical debt term throughout the project life-cycle in all artifacts and processes. While country- and project-specific differences were identified in practitioners' responses in this study, practitioners generally agreed that technical debt must be managed [10,12,40,41]. Disregarding this fact can reduce a project's chances of success and being maintainable in a highly competitive industry.

Finally, we anticipate three threads of research to follow on from the study presented here. First, as we have established a baseline context for technical debt and its management in the agile software development context, it would be useful to identify, classify, and describe technical debt information further for specific practices and as projects evolve. For example, what technical debt information is associated with the most controversial *On-Site customer* practice, and how does the information change when the project advances from an initial phase (e.g., volatile requirements gathering) to a later phase (e.g., meticulous testing and release)? We expect answers to such questions to provide us with an interface through which to integrate separate, explicit, technical debt management procedures. Second, it would be useful to validate, identify, and understand the potential effect(s) of any cultural differences on the outcomes observed in this study. Such insights would help us to provide more granular recommendations for the different territories, in informing their software development practices. Such an inquiry would also benefit from larger country-specific data sets. Third, as the agile techniques, practices and processes discussed here are disconnected from their respective agile methods, we believe that there is an avenue for future work to also correlate and model the synergies between the techniques, their adoption levels, and their respective methods.

### 5.3. Limitations

This study provides an extension to the technical debt knowledge base, however, we concede that the work is also subject to limitations and threats to validity. The framework established by Stavru [36] allows for the systematic assessment of the study's thoroughness and trustworthiness. Thoroughness considers survey definition, design, implementation, execution, and analysis. Regarding survey definition, Sections 3.1 and 3.3 clearly describe this survey's *Objectives* and the *Survey method* respectively.

For the survey's design, our early justifications and initial position in Section 3.2 justify our approach to the design of the survey to capture the required evidence and insights into technical debt. In our conceptualizing of the study we did not account for expected relationships between captured variables, as the survey is exploratory in nature. Hence, there remains a possibility that some overlooked relationships affected the design. Section 3.3 has described the *Data collection method* whilst taking into account *Provisions for securing trustworthiness* (e.g., following established survey design principles and revisions of related work in Sections 3.2 and 2 respectively). The *Target population* for the survey is captured in Section 3.3, but, as we targeted a large sample population with a self-administrated online questionnaire, the *Sample frame* cannot be accurately described. That said, all the available software engineering (development) organization registries were exhausted as described in Section 3.3. Hence, while our results indicated that we were able to contact a broad spectrum of software organizations and practitioners, and we documented accurate Sample sizes for all the three data sets, our *Sampling method* is a *limited random sample*. However, existence of three separate datasets is likely to reduce the effects of sampling bias. We also expect further expansion of the survey to provide additional datasets for consideration.

Regarding the survey's implementation, Section 3.3 addressed *Questionnaire evaluation* (via piloting in Section 3.3) while the *Questionnaire itself* is made accessible online at http://soft.utu.fi/tds16/ questionnaire.pdf. For the survey's execution, the *Media* and the number of *Responses* were documented in Section 3.3. The *Response burden* was communicated to the respondents (see e.g., average response time and confidentiality guarantee in the Questionnaire (see

http://soft.utu.fi/tds16/questionnaire.pdf), and *Execution time* was limited in Section 3.3. *Follow-up procedures* included email remainders, although this had limited effect in cases where personal contact details were not available. No incentives were used to encourage response nor to counter non-response; the survey allowed respondents to give their contact information disconnected from their responses so that we could inform them about future publications. Finally, in direct connection to the *Sampling frame*—and the bias discussed for it—we could only provide partial estimates for the survey *Response rate*. This poses a threat to the generalizability of our results, and to remedy this, the work presents thorough background details analysis in Section 4.1.

Regarding the survey data analysis, its *Assessment of trustworthiness* is affected by the sample frame and response rate issues. In particular, the sample frame error is difficult to account for, given the nature of anonymous web-based surveys. Formal statistical methods are applied wherein possible to enhance trustworthiness (see Section 4), and piloting was used as an informal mechanism to reduce measurement error. *Limitations* of the analysis are addressed here via Stavru's framework [36].

Trustworthiness considers internal validity, external validity, consistency, and neutrality [42]. Regarding internal validity, in general, the survey instrument was designed so that the questions posed were straightforward (e.g., the Questionnaire http://soft.utu.fi/tds16/questionnaire.pdf , Q14: *"Does your team apply Pair-programming?"*), and hence, subject to as little misinterpretation as possible. However, issues still remain in this study that affect its internal validity. First, regarding RQ2 (see Section 3.1) and the associated results (see Section 4.3), we note that there is a possibility that respondents are prone to choosing software components developed by others for criticism. We assume this as it can be easier to pass judgment or to perceive there to be technical debt in components that you have not created. Hence, the recorded concrete technical debt instances may be biased towards software components developed by others, which affects how representative our sample is. A potential negative outcome of this could be that respondents recorded a higher number of legacy origins, given that such systems are often inherited from others. Second, affecting RQ2 as well, we asked the respondents to identify and consider one technical debt instance for the duration of the survey. This means that the recorded instance was based on the subjective decisions of the respondents. This may bias the sample, for example, describing the most prominent or cumbersome technical debt instances the respondents are facing; leaving out or limiting assessments of others. However, capturing these instances from respondents with multifold backgrounds somewhat alleviates these threats to an extent as they indicate the instances to be universally prominent to software engineering.

Section 3.2 relates the study's design to the research questions and Section 3.3 discussed piloting to verify the design (e.g., overcoming language issues discussed in Section 3.3). For external validity, we identified sample frame and response rate to affect generalizability. However, the extensive analysis of the background characteristics in Section 4.1 should limit the challenges to generalization. For consistency, we have described the study design and execution in Section 3 whilst the survey instrument itself is accessible online at http://soft.utu.fi/tds16/questionnaire.pdf. Finally, to increase the study's neutrality we have made an effort to contrast our results against those in related works in Section 5 whilst openly discussing the study's objectives, methods, and limitations in Sections 3.1, 3.2, and 5.3 respectively.

Finally, three potential issues may affect the construct validity of this study. First, while misinterpretation of the questionnaire was likely to be reduced with unambiguous wording, it is possible that, in part, the questionnaire's terminology was difficult to comprehend. For example, discussing the principal and interest aspects of technical debt relied on the respondent's existing knowledge and comprehension of the provided technical debt definition. As neither the 'principal' nor the 'interest' term have their base in the software development domain, an exhaustive definition requires that respondents extend this with knowledge from the financial domain. As most software practitioners cannot be expected to be experts of the financial domain, and the operationalization of the concepts in the provided definitions is minimal, there can be deficiencies in their understanding. This can influence, for example, how the respondent perceived the queried agile software development practices and processes to affect technical debt.

A similar issue can be observed to affect the queried software development characteristics (e.g., transparency and agility). Multiple definitions for these concepts may exist, and the survey instrument did not provide any. However, we trust that the experienced developers at least share common views of these characteristics. As such, we discuss correlations between the characteristics, but we do not link these results to other analysis. Second, it is also possible that respondents' interpreted questions or answer options differently from the intended interpretation. However, the pilot survey did not reveal any significant ambiguities. Third, in Section 4.2.1 we performed statistical analysis to probe if respondents' existing technical debt knowledge correlated with what they indicated as the technical debt effect of particular agile software development practices and processes. We found in Table 2 that the correlation was moderate for *Coding standards* and *Continuous integration*. This indicates that, on these accounts, the existing knowledge of respondents is not independent of the perceived effects and thus threatens validity. However, we noted that the correlation was between existing knowledge and non-positive effects. The correlation here means that if a respondent had less existing knowledge she/he would indicate more negative effects for the technique in question or if she/he had good existing knowledge she/he would indicate no effect for the technique. Considering this bias when interpreting Fig. 7a to answer RQ3, both the *Coding standards* and *Continuous integration* practices have a highly positive effect on technical debt, which means that without the bias this effect could be higher. We are thus satisfied that the pattern of these outcomes are reliable.

# 6. CONCLUSION

This article has reported the design and execution of a software practitioner survey in order to establish the breadth of practitioners' knowledge about technical debt, how technical debt is manifested in projects, and the perceived effects of common agile software development practices and processes on technical debt. It is concluded that while practitioners are aware of the concept of technical debt, this knowledge is implicit, and hence, underutilized (RQ1a). Noting that recent surveys [5,19,21] found technical debt to be a universal issue, this is a cause for concern. We observe that there were no major differences in conformance to the technical debt definitions introduced between different stakeholder groups, which indicates that technical debt can contribute to bridging the communication gap between different software stakeholder groups and roles (RQ1a). A natural next step from this result would be to look for ways in which technical debt-assisted discussions could be facilitated between the stakeholder groups. We also argue that there is a possibility of further exploring the specific role agile software development practices and processes (e.g., retrospectives) play in helping to mitigate technical debt.

We observed that it is difficult to identify concrete instances of technical debt. Based on this, we argued that such instances would be challenging subjects for technical debt management. More closely, a single technical debt instance was seen to span multiple software development phases (RQ2b) whilst being emergent due to a number of causes (RQ2a). This poses an additional challenge for practices and tools which intend to govern these instances: a similar level of agility is required from the tool if its intention is to be able to track an instance through multiple development phases which may concern several software development components for which the stakeholders are completely independent from one another.

Notwithstanding potential bias, for the majority of the recorded technical debt instances, the indicated origins of the instances were associated with software project legacy. Further validation is required for this result, as if this relationship exists, then there is a clear motive to further look, for example, into adopting practices from the more matured software legacy domain to enhance technical debt management. Another implication we discussed from this relationship was the potential for misuse where issues previously considered legacy are rebranded technical debt. By definition the concepts share a commonality, but, arguably, additional measures are required to convert legacy software—which is usually a large sets of software components in various states of disrepair—into manageable technical debt instances [43].

Finally, our analysis showed that most agile software development practices and processes were seen to have—either positive or more diverged—an effect on technical debt and its management (RQ3a). Additional measures are required to identify and validate the implicit or explicit factors in these techniques that contribute towards technical debt management. Our findings here establish the interface through which technical debt management can be enhanced for existing software projects that have established development practices and process in use without introducing new and disruptive, separate technical debt management methods. To that end, techniques safeguarding the structure and state of the software implementation were perceived to have the most positive effect on technical debt (RQ3b). This is reassuring, as most technical debt instances were recorded to reside in the implementation phase of the software development lifecycle. Furthermore, our outcomes also somewhat support the competing stakeholder interest phenomenon discovered in related work as highest opinion diversion regarding effects on technical debt were recorded for agile practices that bring together multiple roles and stakeholder groups. As several software development respondents indicated that they were applying this method, and it is indicated to be crucial for particular agile methods [26], from the perspective of technical debt management, these practices are especially interesting. We believe there is an indicated pattern of a conflict of interests between respondents. If further studies are capable of identifying and resolving these conflict there should be an immediate positive effect on agile practices' perceived technical debt effects.

## ACKNOWLEDGEMENTS

Johannes Holvitie is partially supported by grants from Nokia Foundation, Ulla Tuominen Foundation, and Finnish Foundation for Technology Promotion. Dr. Rodrigo Spínola is partially supported by CNPq Universal 2014 grant 458261/2014-9. The authors would like to thank the reviewers for their extensive comments which have significantly contributed towards improving the quality of this article.

## REFERENCES

[1] K. Beck, M. Beedle, A. Van Bennekum, A. Cockburn, W. Cunningham, M. Fowler, J. Grenning, J. Highsmith, A. Hunt, R. Jeffries, et al., Manifesto for agile software development, 2001, Online, URL: http://agilemanifesto.org/.

[2] K. Power, Definition of ready: an experience report from teams at Cisco, International Conference on Agile Software Development, Springer, 2014, pp. 312–319.

[3] W. Cunningham, The WyCash portfolio management system, Addendum to the Proceedings on Object-Oriented Programming Systems, Languages, and Applications (OOPSLA), (1992), pp. 29–30.

[4] C. Seaman, Y. Guo, C. Izurieta, Y. Cai, N. Zazworka, F. Shull, A. Vetrò, Using technical debt data in decision making: potential decision approaches, Managing Technical Debt (MTD), 2012 Third International Workshop on, IEEE, 2012, pp. 45–48.

[5] Z. Li, P. Avgeriou, P. Liang, A systematic mapping study on technical debt and its management, J. Syst. Softw. 101 (2015) 193–220.

[6] J. Holvitie, V. Leppanen, S. Hyrynsalmi, Technical debt and the effect of agile software development practices on it-an industry practitioner survey, Managing Technical Debt (MTD), 2014 Sixth International Workshop on, IEEE, 2014, pp. 35–42.

[7] S. McConnell, Technical debt, 2007, Online, URL: http://www.construx.com/10x_Software_Development/Technical_Debt/.

[8] P. Kruchten, R.L. Nord, I. Ozkaya, Technical debt: from metaphor to theory and practice. IEEE Softw. 29 (6) (2012) 18–21.

[9] N.S. Alves, L.F. Ribeiro, V. Caires, T.S. Mendes, R.O. Spínola, Towards an ontology of terms on technical debt, Managing Technical Debt (MTD), 2014 Sixth International Workshop on, IEEE, 2014, pp. 1–7.

[10] N. Brown, Y. Cai, Y. Guo, R. Kazman, M. Kim, P. Kruchten, E. Lim, A. MacCormack, R. Nord, I. Ozkaya, et al., Managing technical debt in software-reliant systems, Proceedings of the FSE/SDP Workshop on Future of Software Engineering Research, ACM, 2010, pp. 47–52.

[11] C. Seaman, Y. Guo, Measuring and monitoring technical debt, Adv. Comput. 82 (2011) 25–46.
[12] E. Tom, A. Aurum, R. Vidgen, An exploration of technical debt, J. Syst. Softw. 86 (6) (2013) 1498–1516.
[13] C. Izurieta, A. Vetrò, N. Zazworka, Y. Cai, C. Seaman, F. Shull, Organizing the technical debt landscape, Managing Technical Debt (MTD), 2012 Third International Workshop on, IEEE, 2012, pp. 23–26.
[14] T. Klinger, P. Tarr, P. Wagstrom, C. Williams, An enterprise perspective on technical debt, Proceedings of the 2nd Workshop on Managing Technical Debt, ACM, 2011, pp. 35–38.
[15] W. Snipes, B. Robinson, Y. Guo, C. Seaman, Defining the decision factors for managing defects: a technical debt perspective, Managing Technical Debt (MTD), 2012 Third International Workshop on, IEEE, 2012, pp. 54–60.
[16] A. Martini, J. Bosch, Towards prioritizing architecture technical debt: information needs of architects and product owners, 2015 41st Euromicro Conference on Software Engineering and Advanced Applications, IEEE, 2015, pp. 422–429.
[17] A. Martini, J. Bosch, M. Chaudron, Investigating architectural technical debt accumulation and refactoring over time: a multiple-case study, Inf. Softw. Technol. 67 (2015) 237–253.
[18] R.O. Spinola, A. Vetro, N. Zazworka, C. Seaman, F. Shull, Investigating technical debt folklore: shedding some light on technical debt opinion, Managing Technical Debt (MTD), 2013 4th International Workshop on, IEEE, 2013, pp. 1–7.
[19] E. Lim, N. Taksande, C. Seaman, A balancing act: what software practitioners have to say about technical debt, IEEE Softw. 29 (6) (2012) 22–27.
[20] Z. Codabux, B. Williams, Managing technical debt: an industrial case study, Managing Technical Debt (MTD), 2013 4th International Workshop on, IEEE, 2013, pp. 8–15.
[21] N.A. Ernst, S. Bellomo, I. Ozkaya, R.L. Nord, I. Gorton, Measure it? Manage it? Ignore it? Software practitioners and technical debt, Proceedings of the 2015 10th Joint Meeting on Foundations of Software Engineering, ACM, 2015, pp. 50–60.
[22] W.N. Behutiye, P. Rodríguez, M. Oivo, A. Tosun, Analyzing the concept of technical debt in the context of agile software development: a systematic literature review, Inf. Softw. Technol. 82 (2017) 139–158.
[23] K. Schwaber, M. Beedle, Agile software development with Scrum, 18 Prentice Hall PTR Upper Saddle River, NJ, 2002.
[24] T. Dingsøyr, S. Nerur, V. Balijepally, N.B. Moe, A decade of agile methodologies: towards explaining agile software development, J. Syst. Softw. 85 (6) (2012) 1213–1221.
[25] T. Dybå, T. Dingsøyr, Empirical studies of agile software development: a systematic review, Inf. Softw. Technol. 50 (9) (2008) 833–859.
[26] K. Beck, C. Andres, Extreme Programming Explained: Embrace Change, Addison-Wesley Professional, 2004.
[27] N. Kurapati, V.S.C. Manyam, K. Petersen, Agile software development practice adoption survey, Agile processes in software engineering and extreme programming, Springer, 2012, pp. 16–30.
[28] D. West, T. Grant, Agile development: mainstream adoption has changed agility, Technical Report, Forrester Research, 2010.
[29] P. Abrahamsson, O. Salo, J. Ronkainen, J. Warsta, Agile software development methods: review and analysis, Technical Report, VTT Finland, 2002.
[30] O. Salo, P. Abrahamsson, Agile methods in European embedded software development organisations: a survey on the actual use and usefulness of extreme programming and scrum, IET Softw. 2 (1) (2008) 58–64.
[31] B. Bruegge, A.H. Dutoit, Object-Oriented Software Engineering Using UML, Patterns and Java, Prentice Hall, 2004.
[32] P.L. Alreck, R.B. Settle, The Survey Research Handbook, 2 Irwin Homewood, IL, 1985.
[33] P.V. Marsden, J.D. Wright, Handbook of Survey Research, Emerald Group Publishing, 2010.
[34] Y. Guo, C. Seaman, A portfolio approach to technical debt management, Proceedings of the 2nd Workshop on Managing Technical Debt, ACM, 2011, pp. 31–34.
[35] P.M. Nardi, Doing Survey Research: A Guide to Quantitative Methods, Pearson Allyn & Bacon, 2006.
[36] S. Stavru, A critical examination of recent industrial surveys on agile method usage, J. Syst. Softw. 94 (2014) 87–97.
[37] J. Cohen, A power primer. Psychol. Bull. 112 (1) (1992) 155.
[38] H. Kagdi, M.L. Collard, J.I. Maletic, A survey and taxonomy of approaches for mining software repositories in the context of software evolution, J. Softw. Maint. Evol.: Res. Pract. 19 (2) (2007) 77–131.
[39] VersionOneInc., 9th Annual State of Agile Survey, Technical Report, VersionOne, Inc., 2015. URL: http://www.stateofagile.com/.
[40] F. Buschmann, To pay or not to pay technical debt, IEEE Softw. 28 (6) (2011) 29–31.
[41] Y. Guo, C. Seaman, R. Gomes, A. Cavalcanti, G. Tonin, F. Da Silva, A. Santos, C. Siebra, Tracking technical debt - an exploratory case study, Software Maintenance (ICSM), 2011 27th IEEE International Conference on, IEEE, 2011, pp. 528–531.
[42] E.G. Guba, Criteria for assessing the trustworthiness of naturalistic inquiries, Edu. Technol. Res. Develop. 29 (2) (1981) 75–91.
[43] J. Holvitie, S.A. Licorish, A. Martini, V. Leppänen, Co-existence of the technical debt and software legacy concepts, Technical Debt Analytics, 2016 1st International Workshop on, CEUR-WS, 2016, pp. 80–83.